# Comprehensive Review and New Analysis Software for Single-file Pedestrian Experiments

Rudina Subaih[1,2] · Antoine Tordeux[2] · Mohcine Chraibi[1]

[1] Institute for Advanced Simulation, Forschungszentrum Jülich, Jülich, Germany,
E-mail: r.subaih@fz-juelich.de, m.chraibi@fz-juelich.de
[2] School for Mechanical Engineering and Safety Engineering, University of Wuppertal, Wuppertal, Germany,
E-mail: rudina.subaih@uni-wuppertal.de, tordeux@uni-wuppertal.de

**Abstract** This paper offers a comprehensive examination of single-file experiments within the field of pedestrian dynamics, providing a review from both theoretical and analytical perspectives. It begins by tracing the historical context of single-file movement studies in pedestrian dynamics. The significance of understanding the fundamental relationships between density, speed, and flow in pedestrian dynamics is explored through the lens of simple single-file systems. Furthermore, we examine various traffic systems involving human or non-human entities such as ants, mice, bicycles, and cars, and provide insights. We explore the types of experimental setups, data collection methods, and factors that influence pedestrian movement. We also define and explain the common concepts related to single-file movement, particularly in experimental research. Finally, we present a Python tool named "SingleFileMovementAnalysis" designed for analyzing single-file experimental data, specifically head trajectories. This tool provides a unified approach for computing movement metrics like speed, density, and headway. The article aims to stimulate further research and underscore the areas where future researchers can contribute to the advancement and improvement of single-file studies.

# 1. Introduction

In their seminal work, Seyfried et al. [1] present the concept of single-file movement in pedestrian dynamics to explore the relationship between density, flow, and mean velocity, also known as the fundamental diagrams, within pedestrian traffic. The fundamental diagram quantifies the capacity of pedestrian facilities, allowing the assessment of escape routes and the evaluation of pedestrian models. To assess dependence on the fundamental diagram, Seyfried et al. investigate experiments of single-file movement, where pedestrians walk unidirectionally along a line with reduced degrees of freedom. This restricts the possible factors that influence the fundamental diagram. In 2009, Chattaraj et al. [2] replicated the same experiment in India [1], with the main aim of analyzing the cultural influence (social conventions) on pedestrians' movement. The motivation behind performing single-file experiments, as pointed out by Chattaraj et al., is that the density-speed relation is influenced by multiple factors that are still not completely understood. In general, the importance of studying single-file movement can be traced back to the open questions: Which factors influence the fundamental relationships? What are the possible movement quantities that describe the walking characteristics of pedestrians?

Over the past decade, several experiments have been conducted to explore single-file movement. The objective of these experiments is to identify basic relationships within a system using a minimal number of variables and parameters. In these experiments, researchers typically set up a controlled environment in which pedestrians are asked to walk through a narrow corridor without overtaking. Figures 1(a) and 1(b) show the publication trends over the years and countries/territories, respectively. The surge in publications in

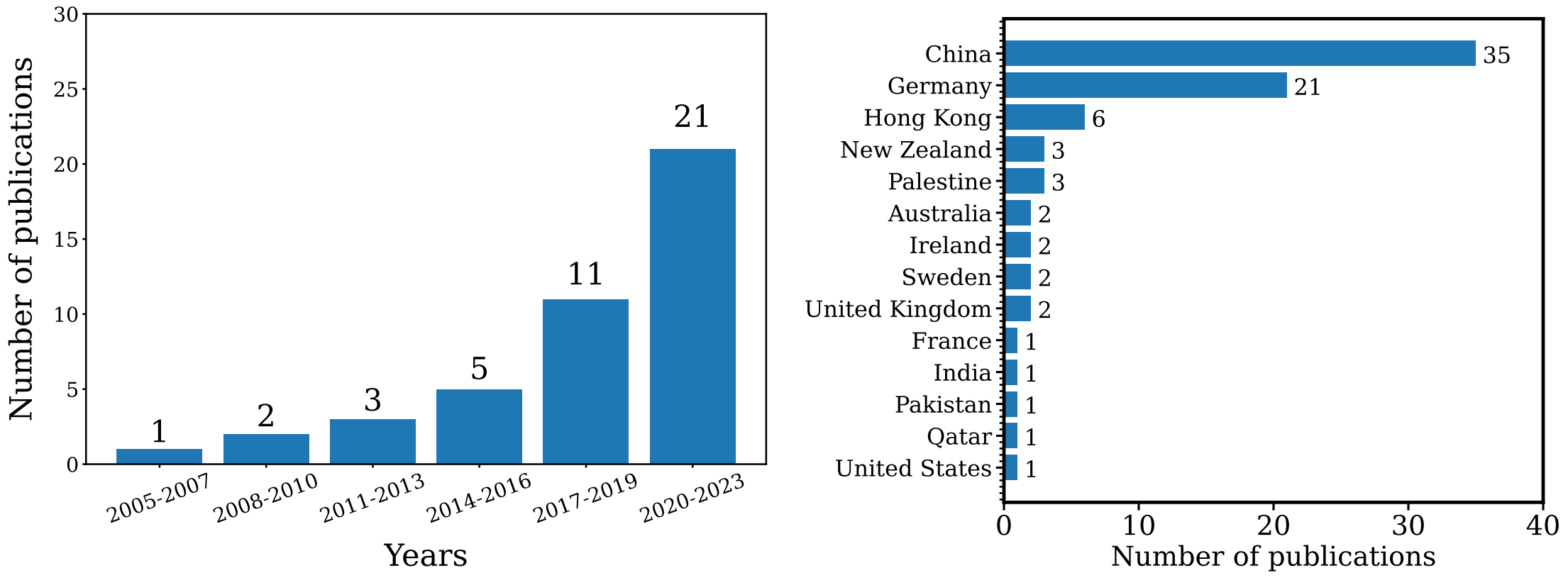

(a) Publications by year.

(b) Publications by country/territory (based on the affiliation of the authors).

**Figure 1** The number of publications that mentions *single-file movement pedestrian dynamics* or *single-file motion pedestrian dynamics*, according to a Scopus search on 29 March 2024.

recent years shows a rising interest in single-file movement in pedestrian dynamics. However, it is worth noting that the terminology *single-file movement pedestrian dynamics* or *single-file motion pedestrian dynamics* is a relatively recent concept that, until now, has not been well-established (see the number of publications in Figure 1(a)). We can divide the research focus of publications on single-file movement in pedestrian dynamics into four main topics: experiments, data analysis, modeling, and experiments with models (see Figure 2).

Given the importance of single-file experimental research, conducting a comprehensive literature review is essential to identify the gaps in previous studies and outline directions for future research. Xue et al. [3] examine and compare pedestrian single-file experiments from a modeling perspective. They compare the basic characteristics of pedestrian movement in the literature. Their work covers methods for measurement, data extraction, stepping behavior quantities, influential factors, and simulations of single-file pedestrian flow. Still, a more in-depth review, focusing on the details of the experiments from a data analytical viewpoint, is required. In this work, we explore various traffic systems, including humans, mice, ants, bicycles, and cars, to identify similarities and differences that can improve pedestrian dynamics. Furthermore, we define different pedestrian single-file systems and discuss their types. We characterize the types of experimental setups and identify factors that influence movement, along with discussion. Moreover, we propose a methodology for preparing trajectory data and calculating movement quantities using an open-source Python tool called "SingleFileMovementAnalysis" [4], which is essential for enabling future research to build on.

The subsequent sections of this paper are structured as follows. In Section 2, we explore the single-file traffic systems available in the literature and provide comparative insights. Additionally, we characterize single-file pedestrian systems. In Section 3, we review the single-file experiments in the literature focusing on the type of setups. In Sections 4, we explore the data collection methods adopted and the movement quantities investigated in the single-file experimental research. In Section 6, the factors influencing pedestrian movement are identified and studied. In Section 7, we propose a methodology for preparing trajectory data, computing in a systematic way movement quantities and present a Python software tool to analyze single-file movement data. Finally, in Sections 8 and 9, we provide a summary of the findings, identify trends and open issues, and suggest future research directions.

## 2. Exploring Single-File Traffic Systems: Definition and Comparative Insights

Several single-file experiments have been conducted to investigate human movement [1, 2, 5–41]. After reviewing the literature above, we define the *single-file pedestrian system*, following the general definition of a system as described by Backlund et al. [42], as a group of interacting pedestrians walking in a narrow path (physical or virtual path [22]), where individuals cannot pass each other (rule: no overtaking). The order of the pedestrians remains constant throughout the experiment. In this context, the system aims to question the basic elements of pedestrian movement, including physical and psychological interactions.

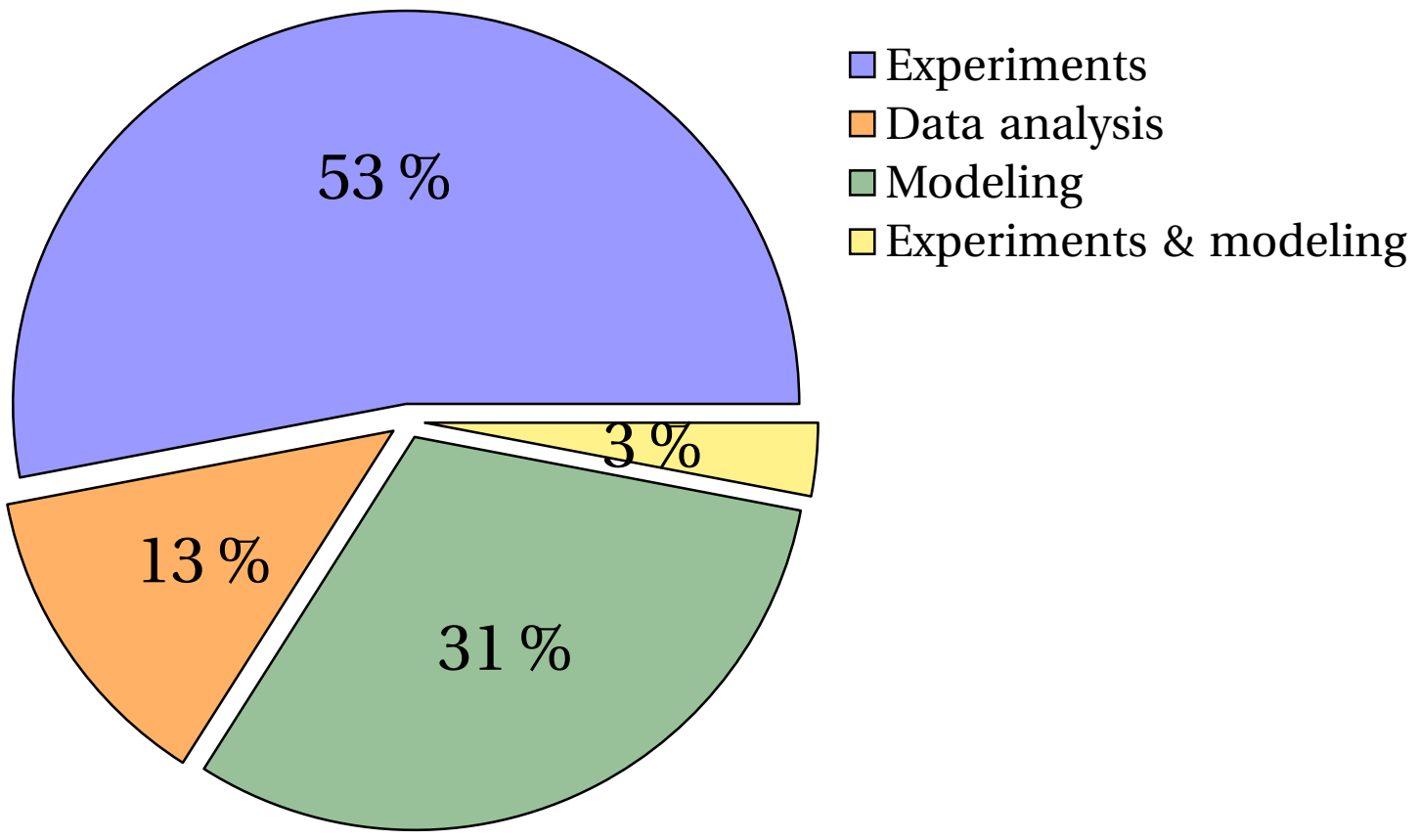

**Figure 2** The percentage distribution of single-file movement publications in pedestrian dynamics across various subjects, from the literature reviewed for this paper.

The single-file system can be a *closed system* or an *open system*. In a closed system, pedestrian movement is influenced by elements within the system. Whereas in an open system, the surroundings can influence pedestrian movement. The term "surroundings" refers to the environment adjacent to the area of interest. For example, when pedestrians leave the predefined system boundaries and interact with the external environment. Further explanation of the open and closed single-file systems is described in Section 3. Having defined the pedestrian single-file system, this section aims to identify similarities between human and non-human single-file systems. We examine the basic principles of movement that govern these systems and identify possible movement similarities.

Exploring other single-file systems involving non-human entities offers valuable insights into understanding movement properties and relations in these systems. For example, studying the adaptive behaviors of ants and mice, and observing the movement of bicycles, and cars in response to movement stimuli (obstacles, other nearby entities, etc.) can inspire innovative modeling or crowd management approaches. Table 3 in Appendix A summarizes all single-file experiments reviewed in this article for various traffic systems.

Many non-human single-file systems, such as those observed in insects and rodents within animal societies, have been explored in the literature [43, 44]. Both systems (mice and ants) show that speed decreases with increasing density and exhibit a piecewise linear relationship between headway distance and speed, similar to the human system. However, scattered data points are observed in these relationships. The researchers attribute this primarily to random pauses. For example, Xiao et al. [43] find that at all densities, mice stop under various circumstances, including spontaneous pauses, space constraints, and tail effects (when a mouse stops or retreats after being touched by the tail of another). Similarly, Wang et al. [44] observe that ants exhibit random pauses during their experiments. Unlike in human systems, stopping occurs at high densities only when insufficient space is available to move forward [45].

Another difference is that mice and ants do not maintain personal space while walking,

resulting in increased speed and flow at high densities. For instance, in the experiment with mice, the flow remains almost constant at high densities (non-dimensional density above 0.4) because the mice tend to make contact and move on top of each other, a behavior we refer to as overlapping. Like in experiments with ants, behaviors such as touching and moving backward are observed. Unlike the human system, where flow and speed decrease at high densities because pedestrians maintain some distance to avoid collisions and touching others. We recognize that differences in movement can be attributed to the dissimilar physical attributes (i.e., body size and shape), cognition, and decision-making processes of humans and non-human beings. However, we assume that touching and pausing behavior helps to gain insight into improving flow in high densities (short headway distances less than personal space).

Another group of single-file systems studied in the literature is vehicular systems. Research on vehicular single-file movement shows good agreement between studies regarding the relationship between certain movement quantities [46], such as the density-flow and density-speed. However, vehicles such as bicycles [14,46,47] and both human-driven and autonomous cars [48–51], are machines controlled by humans. This indicates that the movement of these vehicles is systematic and dominated by the physical constraints on the car, such as inertia and limitations on possible acceleration. We assume that investigating vehicular systems helps us understand how humans make decisions to control vehicles, addressing three main concerns: following instructions, avoiding collisions, and ensuring safety. Thus, the benefits of studying pedestrian dynamics from studying vehicular traffic can be linked to understanding cognitive processes. The differences and similarities in the motion properties among single-file traffic systems (such as pedestrians, mice, ants, bicycles, and cars) are summarized in Table 1.

## 3. Types of Experimental Setups

This section reviews the setup configurations and discusses their distinct features of single-file experiments involving pedestrians. We also present previously studied setup types in the literature and provide some insights.

Experimental studies on pedestrians' single-file movement have been performed in various shapes/types of setups (see Figure 3): oval [1, 2, 5, 6, 8, 11–13, 17, 19, 21, 25, 26, 28, 29, 32–34, 38, 39, 41], circle [7, 9, 10, 14, 16, 22, 24, 32], stairs [15, 35, 36], one-dimensional observation area [22, 27, 31], square with four straight corridors and four arcs [40], rectangle [30], rectangle with four straight corridors and four arcs [23], ship corridor [20], branch [37], seat aisle [18], flood [52].

We observe that the selection of the shape/type of the experimental setup is contingent upon the *evacuation scenario* the authors intend to investigate. After reviewing the literature, we categorize single-file experiments into five evacuation scenarios based on the evacuation facility under study: flood (moving in water), stairs, ships, seat aisles, and ground level (in general).

Here, we provide a brief overview of the relevant literature on the evacuation scenario in flood, stairs, ship corridors, and seat aisles. Li et al. [52] investigate the effectiveness of different formations for evacuating pedestrians during a flood. The authors perform experiments with a pool, using a single-file system at two specific water depths (0.35 m and 0.60 m), and compare the efficiency of evacuations with and without a rescue rope. The study finds that using a rescue rope in single-file formation during flood evacuations significantly reduces pedestrian fatigue and increases speed, particularly in higher water depths. In the investigation of stair evacuation, Chen et al. [15] conducted experiments exploring the movement characteristics of pedestrians ascending and descending stairways. The results show that descending stairs is faster than ascending, as pedestrians benefit from gravity during descent, whereas ascent requires more effort, resulting in slower speeds. Furthermore, the speed in stairways described by the number of steps in the longitudinal direction. Wang et al. [35] further investigate the impact of stair configuration and explore the influence of stair dimensions on pedestrian movement characteristics. The authors find that the stair configuration, particularly tread depth and riser height, signifi-cantly affects pedestrian movement speed, with steeper stairs leading to reduced walking speed. Ye et al. [36] compare pedestrian movement under motivation (fast walking) with normal walking. The results show that pedestrians on stairs move faster when motivated (fast walking condition), with descending movements being quicker than ascending ones, and that motivation increases velocity correlation between adjacent pedestrians.

**Table 1** Comparison of movement characteristics among different single-file traffic systems.

| Traffic system | Keep distance in front | Sensitivity to distance in front in controlling the speed | Overlap behavior | Pauses/stopping behavior | Backward movement |
|---|---|---|---|---|---|
| **Human** | Yes, respect personal space | Sensitive | Does not occur | Stop-and-go waves at high densities | Rarely (when someone unintentionally collides with the proceeding) |
| **Mice** | No | Not sensitive | Occurs | At all densities (spontaneous pauses because of space constraints, and tail effects) | - |
| **Ants** | No | Not sensitive | Occurs | Short pauses | Occurs (despite the large distance available in front) |
| **Bicycles** | Yes, keep distance to avoid potential collisions | Sensitive | Does not occur | Stop-and-go-waves at high densities | Does not occur |
| **Cars** | Yes, keep distance to avoid potential collisions | Sensitive | Does not occur | Stop-and-go-waves at high densities | Does not occur |

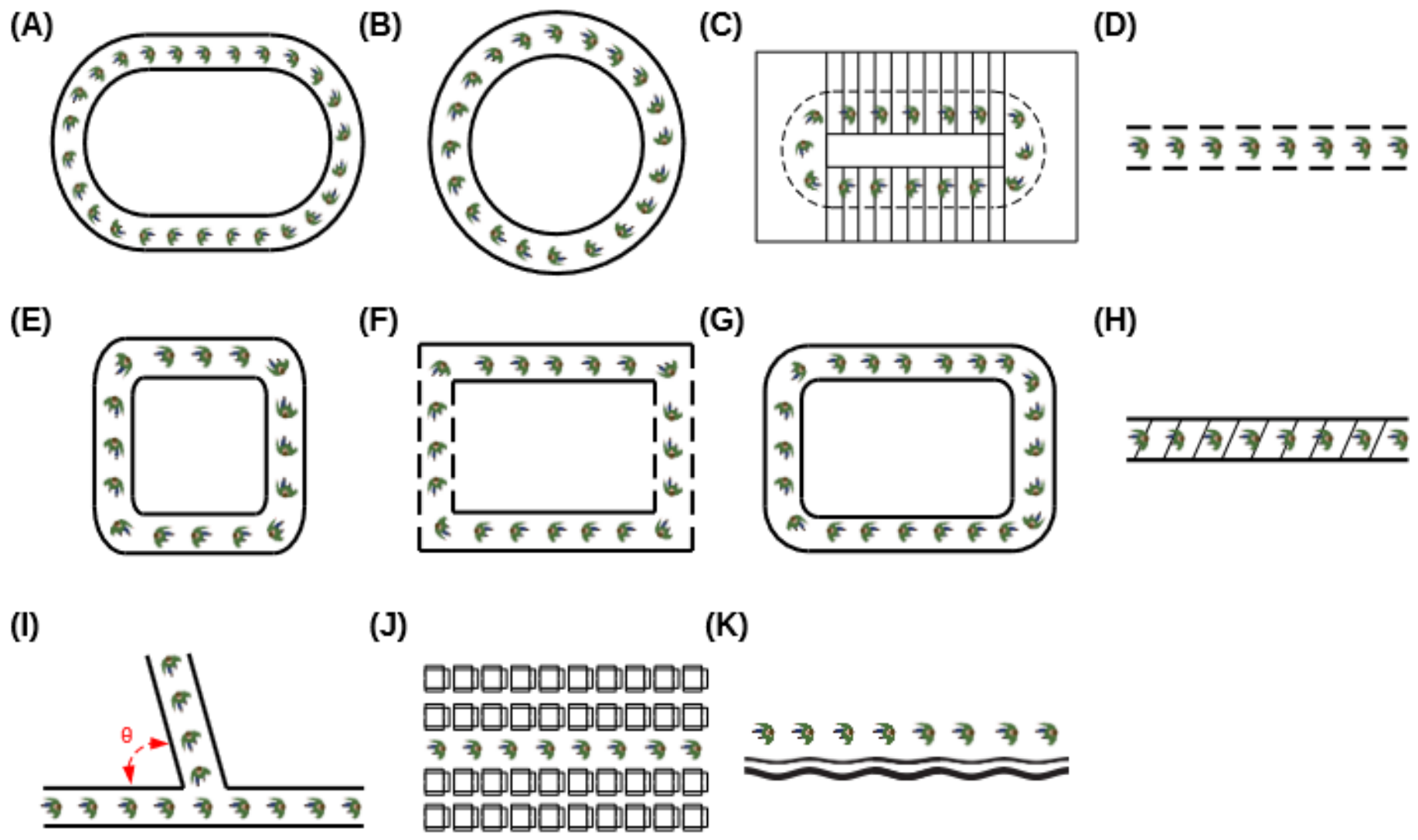

**Figure 3** Illustrations of the experimental setups for different evacuation scenarios: (A) Oval (B) Circle (C) Stairs (D) One-dimensional observation area (E) Square with four straight corridors and four arcs (F) Rectangle (G) Rectangle with four straight corridors and four arcs (H) Ship corridor (I) Branch (J) Seat aisle (K) Flood.

Shifting the focus to evacuation in ship corridors, Sun et al. [20] design a simulator for ship corridors to explore the impact of trim (ship's tilt along its length) and heeling (ship's tilt to one side) on walking characteristics. The results indicate that the trim and heeling angles affect the pedestrian walking speed, with trim angles having a greater impact than heeling. Lastly, for seat aisle evacuations, Huang et al. [18] explore the effects of inactive pedestrians (non-moving), and aisle width's impact on pedestrian dynamics. They find that in narrow seat aisles, pedestrian walking speed increases as aisle width increases up to 0.40 m, after which it stabilizes, and that interactions with inactive pedestrians can significantly slow down the flow, particularly in narrower aisles. While the studies above offer valuable perspectives on single-file movement, our research aims to narrow the focus to ground-level experiments.

In ground-level experiments, various shapes/types of setups are explored. We divide them into two groups depending on the boundary conditions under which the experiment is conducted: *open* (open system) or *closed boundary conditions* (closed system). Experiments under open boundary conditions include setups with open entrances so pedestrians can enter and leave during the experiment. Examples include branch and one-dimensional observation areas. Lian et al. [37] employ a branch setup in which pedestrian streams from two entrances converge into a single main channel to reach the exit. The authors aim to explore pedestrian movement properties through single-file merging experiments, varying merging angles and inflow rates. In the one-dimensional observation area, Appert-Rolland et al. [22] conducted unidirectional experiments to investigate collective and individual decisions in walking. In other words, they study how pedestrians adapt their trajectories and velocities while walking freely in a group of people, rather than moving within a fixed density of pedestrians. During the experiments, pedestrians move along a fixed straight line across the facility, one after the other, following a leader who walks at either their free velocity or a prescribed low velocity.

Huang et al. [27] performed a one-dimensional observation area experiment to analyze the impact of luggage on pedestrian flow at traffic terminals. Participants are instructed to imitate walking in a terminal by following the queue while passing through the observation area. Wang et al. [31] also conduct a one-dimensional observation area experiment to study knee and hand crawling evacuations in fire accidents. Participants pass through a narrow channel divided into two parts: an upright walking area and a knee and crawling area, allowing the investigation of the sole movement characteristics of pedestrians and their movement properties under an increasing inflow at the channel's entrance. In the aforementioned studies, we observe that the authors opted for an open-boundary setup because they are interested in monitoring inflow and outflow as experimental setups.

In experiments under closed boundary conditions, the configuration is enclosed, enabling pedestrians to move within the setup without exiting during the experiment. Examples include an oval, circle, rectangle, a rectangle with four straight corridors and four arcs, and a square with four straight corridors and four arcs. The most commonly explored shape/type is the oval; approximately 52% oval from the total single-file experiments reviewed for this article (for all evacuation scenarios). Seyfried et al. [1] are the first researchers who introduce the oval setup for pedestrian's single-file experiments. The authors explain that the oval setup, similar to the one in [53], limits the number of test objects in the experimental setup and achieves high density without boundary effects. Besides, implementing circular guiding of the passageway gives periodic boundary conditions.

Experiments involving single-file movement in a circle shape or type constitute approximately 19% of the total single-file experiments. The initial research adopting the circle shape in single-file experiments is done by Jezbera et al. [7]. Subsequent studies have continued to perform circle experiments [9, 10, 14, 16, 22, 24, 32]. None of the researchers explicitly state the rationale behind choosing the circle over the oval configuration. Jezbera et al. [7] merely state that they chose a geometry allowing pedestrians to walk in a single line without overtaking, to perform experiments at various pedestrian densities, and to operate in closed boundary conditions. After reviewing the literature in oval and circle shapes, we summarize the main purpose of the experiments as presented in Table 2.

In ground-level experiments under closed boundary conditions, few researchers study single-file movement using the following setup shapes/types: a rectangle, a rectangle with four straight corridors and four arcs, and a square with four straight corridors and four arcs. Wang et al. [30] investigate the movement characteristics of pedestrians during the deceleration phase. The experimental setup employs a rectangular configuration; the rationale behind using a rectangular shape is not explicitly stated. This configuration con-sists of two horizontal and longitudinal paths. The authors emphasize the significance of understanding the deceleration phase in real-life scenarios, where pedestrians slow down to avoid collisions when their predecessors suddenly come to a stop. The focus of Wang et al.'s article is on examining different stop-distance commands: normal stop and close stop, for two types of walking speeds, namely normal and fast walking. Cao et al. [23] investigate the influence of the pedestrian's visibility on the movement properties in a rectangle with four straight corridors and four arcs setup. The authors perform three types of experiments under limited visibility: 0.3% (partial visibility), 0.1% (partial visibility), and 0.0% (no visibility) light transmissions. The shape of the setup has four straight corridors with three arcs built with longitudinal walls. These long walls serve as boundaries to ensure that participants remain within the experimental setup while walking with limited visibility.

**Table 2** Summary of the objectives of oval and circular single-file experiments.

| Main Objective | Focuses | References |
|---|---|---|
| **Investigate movement characteristics or behavior** | Distances between pedestrians | [7] |
| | Density-speed relationship | [1] |
| | Instantaneous velocity and spatial headway Relationship | [10] |
| | Microscopic movement characteristics (density-speed, lateral sway, step frequency, headway distances, and speed-headway distances) | [11] |
| | Stepping behavior (step length, step duration, stepping synchronization, step extent, and contact buffer) | [19, 34, 54] |
| | Movement in high-density conditions | [24] |
| | Influence of bottlenecks on pedestrian flow | [55] |
| **Validate data extraction methods** | Trajectories of pedestrians' heads | [6] |
| **Effect of influential factors** | Rhythm | [9, 16] |
| | Instructions (walking decisions in crowds) | [22] |
| | Social conventions and location | [2, 41] |
| | Age | [12, 25] |
| | Gender | [26, 39, 56] |
| | Background music | [28] |
| | Height constraints | [29] |
| | Social distancing measures | [33] |
| **Compare traffic systems** | Cars vs. bicycles vs. pedestrians | [14] |
| **Compare data sources** | Experiments vs. field studies | [17] |
| **Compare setup shapes** | Oval vs. circle | [32] |

From reviewing the ground-level experiments, we observe that the selection of open or closed shapes/types depends on the goal of limiting the number of pedestrians inside the experimental setup and achieving high density without encountering boundary effects. Additionally, it depends primarily on the *purpose of the experiment*. For example, Lian et al. [37] aims to investigate the effect of complex structures (pedestrians merging on branching walking paths) on the properties of pedestrian movement. Another experiment by Seyfried et al. [1], where they execute an oval setup to analyze the simple system of pedestrians walking at different densities and without boundary effect. However, some researchers do not explicitly state the reason for choosing the shape/type of the experimental setup, but we can deduce it based on the experimental information and details provided.

In summary, we offer valuable insights and recommendations derived from a comprehensive review of the literature on the shapes of setups and experimental settings. We recommend having fewer variables in the experimental settings. That emphasizes isolating undesired effects from the surrounding environment, including external sounds, weather changes, and light changes. Any variation in the experiments can impact the way pedestrians walk. Some research already examines the potential effect of the setup configuration (oval and circle) on pedestrian movement [1, 8, 32]. The oval setup consists of two straight parts and two curvatures, whereas the circular setup is entirely composed of a continuous curve. Seyfried et al. [1] consider the possible influence of the curve part of the oval setup. To avoid this effect, they widen the width of the corridor in the curves, and a measurement section is selected in the center of the straight part of the passageway. However, we assume that limiting the investigation only to the straight part will neglect the characteristics that could be explored in the entire walking path. To avoid the previous issue, Ziemer et al. [13] proposes transforming the oval trajectories into straight trajectories. In this case, the investigation of all trajectories is applicable.

From observing some oval experimental videos, we notice that the navigation between the two parts (straight and curved) could be responsible for a change in walking behavior because the pedestrian turns at the beginning of the curve. The study of [13] already assumes the potential influence and compares the fundamental diagram relationship (density-speed) of the straight and curved parts. They use the Kolmogorov-Smirnov test to determine whether two data sets in the density-speed relationship have the same distribution. The results show that the difference between the straight and curved parts can be neglected.

Fu et al. [32] have another opinion about the possible influence of the curve. The authors examine the impact of curvature by comparing oval and circular pedestrian experiments while keeping settings like path circumference, participant number, methods to extract trajectories, movement direction, and measurement techniques constant. They find that pedestrian flow in the straight part of the oval setup is 20% higher than in the curved part of both setups. This difference is attributed to a more heterogeneous distribution on straight paths, allowing efficient space use and increased flow, whereas curvature leads to a more homogeneous distribution and reduced density. Additionally, at high global densities, the mean instantaneous density is higher in the oval passage than in the circular one. The curvature effect causes differences in pedestrian distribution and decreases density. These findings highlight significant differences in movement characteristics between oval and circular setups. Therefore, we advise researchers studying experiments involving curves to either standardize turning angles for experiments that aim to compare or use experiments with similar shapes.

# 4. Data Collection

This section provides an overview of the data collection processes for pedestrians' single-file experiments conducted under closed-boundary conditions. This section does not explore the devices suitable for data collection in achieving the experiment objectives. However, we provide an overview of the data collection processes in the literature, the data types, and the devices used to collect data from single-file experiments. For more details, we refer to Table 4 in Appendix B

We define *data collection* in single-file experiments as a systematic process for collecting and processing data to investigate the characteristics of pedestrian motion. Several data collection processes are followed depending on the *data type, devices, and methods* used for data collection. The process mainly includes the following steps: installing the devices to collect data (i.e., capturing videos and detecting brain signals) and processing the data (e.g., extracting head positions by detecting pedestrians' heads and tracking them throughout the experiment duration). Based on the experiments we review, the data collection processes can be categorized into two groups:

1. **Semi-automatic data collection**: combines both manual and automatic processes. In other words, some tasks or functions in the data collection processes are automated, while others require human intervention. For instance, Chattaraj et al. [2] use a digital camera to capture the experiments and manually extract the data frames of participants entering/exiting from the measurement area by observing the videos.

2. **Automatic data collection**: all processes are fully automated. The only involvement of humans is to verify and manually adjust the results from the system. For example, Paetzke et al. [39] capture the whole experiment using a digital camera and then detect and extract pedestrians' heads using PeTrack [57] software.

The first step in the data collection involves employing the *appropriate devices* to collect data required for the investigation. In single-file experiments, various devices are installed to collect data and differ in the type of data they measure (see Figure 4).

The primary focus of most experiments is to capture pedestrians' positions over time through *head trajectories* [5,6,11–14,19,21–26,28–30,32,33,39,40,58], which is significant for calculating movement quantities such as speed, density, and headway distances. *Cameras* are the predominant devices used to collect head trajectories. The cameras capture video footage, enabling the extraction of trajectories by detecting and tracking the positions throughout the experiment execution. This results in 2D or 3D positions over time. Various types of cameras are utilized for this purpose. The most commonly used are the *digital cameras* to capture the experiment from a side-view [1, 5, 38] or bird's-eye view [6, 11–14, 19, 21, 23, 25, 28–30, 32, 33, 37, 39, 58]. The former condition (side-view) is recommended when the roof of the experimental hall is not high enough to locate the camera perpendicular to the setup, or if the researchers are interested in observing the movement characteristics from the side view. Whereas, the latter condition (bird's-eye view) provides the data of the overall periodic movement of all pedestrians inside the entire setup. Other types of cameras used rarely in the experiments include *Stereo Vision camera* [1], *UAV drone camera* [24], *infrared camera* [10, 22], and *Camcorders device*. [34].

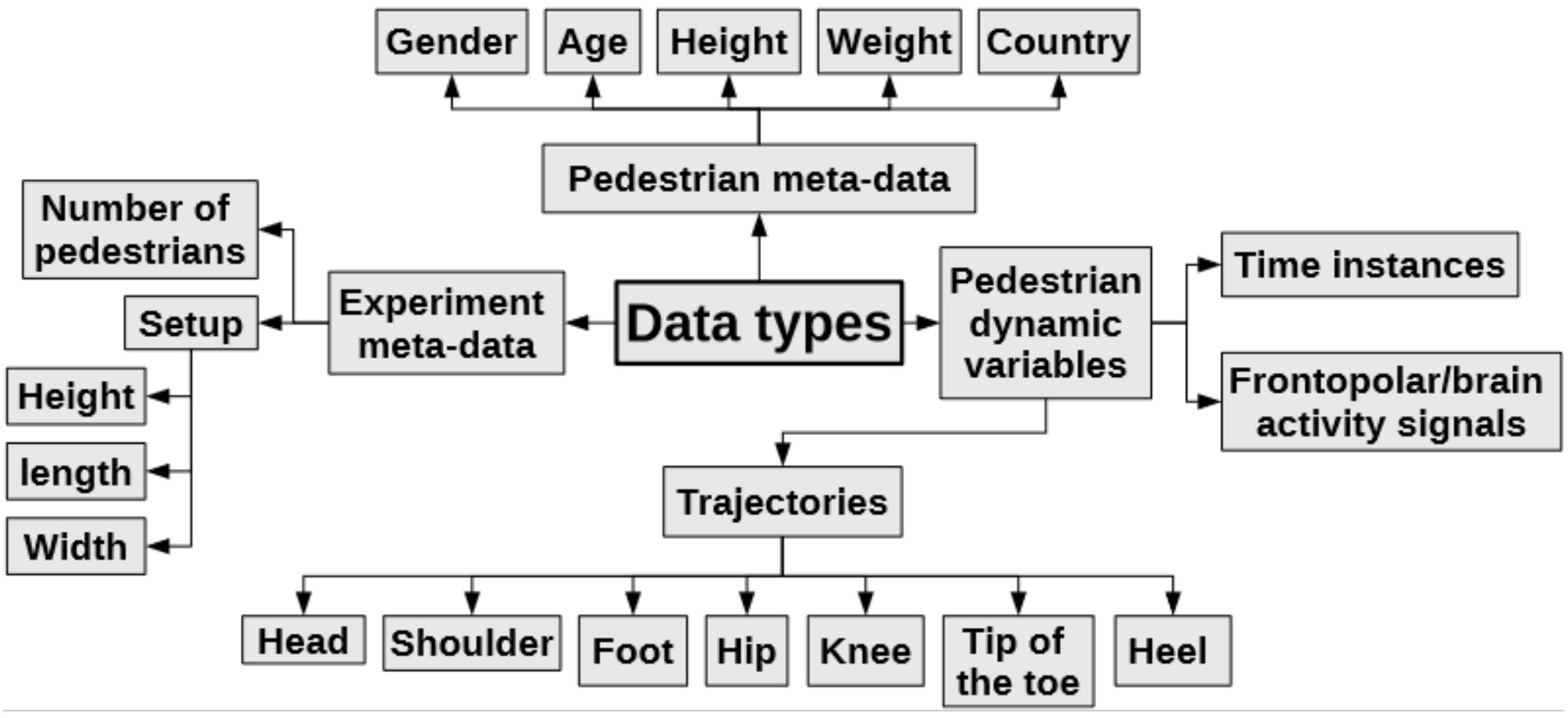

**Figure 4** The data types presented in the single-file movement articles (under closed-boundary conditions) from the literature in this article.

More types of data are extracted from video footage. Thompson et al. [38] collect trajectories of shoulders, *hips*, *knees*, *tips of the toes*, and *heels* to analyze stepping behavior. Furthermore, *time instances* of entry/exit to/from a specified measurement area are recorded to calculate the density in [1, 2, 5, 17, 41]. Other devices are less commonly used in the literature for extracting movement data, such as the *light gate* [7] detect each pedestrian's crossing time at a designated spatial point, ultra-small near-infrared spectroscopy *(NIRS)* device [16] to measure *frontopolar/brain activity signals*, and *Ultra-Wideband (UWB)* to collect pedestrians' trajectories by utilizing tag signals combined with the location coordinates of the base station [40]. After reviewing the literature, we include that the selection of data collection devices depends on the types of data one aims to measure or record to investigate quantities related to movement. Besides, this choice is influenced by the researchers' preferences, which are shaped by the availability of both experience and financial resources to explore and implement new, specialized devices.

The second step for collecting data involves the *processing of the collected data*. It includes extracting the data of interest from collected raw data (i.e., video footage) and preparing the data for usage. One of the most common processing steps for video footage is the extraction of pedestrians' head trajectories over the experimental duration. To achieve this, the process begins by detecting individuals' heads or markers in the initial frame and then tracks their positions in subsequent frames. In addition to the videos, there are other data types, such as pedestrian information stored in an ID marker [39]. Several methods employed in the literature to process the data, such as *manual observation* of the videos [1, 17, 41], applying image processing techniques based on the *mean-shift algorithm* [19,23,33,37], *Tracker software* [24], and *PeTrack software* [5,12,21,25,26,28–30, 32, 39]. PeTrack [57] is the widely used open-source software in the literature because it is specialized software for calibrating, recognizing, and tracking pedestrians and is available online for free. Based on our literature review, we conclude that the data processing varies depending on the utilization of collected data in the investigation (i.e., calculating movement quantities using pedestrian positions).

# 5. Movement Quantities

After collecting the data, the quantities that characterize pedestrian dynamics are calculated. The researchers use these quantities to quantitatively analyze pedestrian dynamics. In this section, we narrow the focus to the research on ground-level experiments conducted under closed boundary conditions. We discuss the quantities and the methodologies employed.

We can categorize the movement quantities in the literature into four groups based on their focus on different aspects of human behavior: quantities to describe *head movement* (to represent pedestrian movement) [1, 2, 6–8, 10, 12–14, 17, 22–26, 29, 30, 32, 33, 39, 41, 55, 56, 58], *stepping locomotion* [9, 19, 21, 34, 38, 54], *both* (head movement, stepping locomotion) [11, 28], and *cognitive behavior* (using brain signals) [16]. Here, we focus the review on the research that analyzes head movement.

Different methodologies are employed in the literature to calculate movement quantities. These methodologies vary according to the objectives of the analyses. The first aspect is the level of movement to describe, including *microscopic* [12, 13, 26, 30, 31, 33, 37, 39] and *macroscopic* levels [1, 2, 28, 41]. At the microscopic level, the movement properties of each pedestrian are investigated during the experiment. At the macroscopic level, the motion characteristics of a group of pedestrians are studied throughout the experiment and averaged over time or space. Jelic et al. [10] qualitatively analyze the influence of different measurement procedures—macroscopic and microscopic—which they refer to as global and local measurements, respectively. Comparing the density–speed diagrams from both measurements reveals very similar results at low densities (approximately less than 1.2 $m^{-1}$). At higher densities (when stop-and-go waves appear), the results of both measurements differ. Ren et al. [25] find that both macroscopic and microscopic level measurements reveal similar trends in density-speed diagrams but with different levels of resolution. The microscopic level measurements provide finer detail, particularly at higher densities, where localized fluctuations in speed and density become pronounced. We observe from reviewing the literature that using macroscopic measurements, where the movement quantities are averaged for multiple pedestrians, ignores the individual movement characteristics. Further quantitative research is needed to compare the disparities in the results from various measurement procedures in single-file experiments. Previous studies show that different measurement procedures produce varying

density–speed relations [59]. However, these findings are based on studies of crowds in straight corridors and T-junction experiments, not on single-file movements.

The second aspect is the setup area that the measurements cover. Studies focus on either the *measurement area* (a predefined part of the experimental setup) [2, 23, 26, 28, 29, 32, 33, 39, 41], or the *entire setup path* (applying a linear transformation or 2D calculations) [12, 13, 25, 28]. Upon reviewing the literature, we notice that calculating movement quantities for a specific part of the setup is simpler. It is simple because there is no need to transform the trajectories when analyzing longitudinal movement (along the x-axis). Instead, the equations for calculating quantities are applied directly to that area. We discuss this further in Section 7. However, analyzing pedestrian movement across the entire setup enables observing phenomena like stop-and-go waves that require complete trajectories [13].

The third aspect is the dimension for calculating movement quantities: *one dimension* or *two dimensions*. Most studies focus on the 1D movement because the researchers are interested in studying the longitudinal interactions among pedestrians walking in single-file experiments. Only Fu et al. [32] calculate the speed and density in 2D in the circle experiments without reporting why they used the 2D measurements. Yet, no single-file research compare the analysis results using 1D and 2D measurements. Using data from Paetzke et al. [39] experiments, we plot the speed-density relation to observe the differences between 1D and 2D measurements (using the tool in Section 7). We disregard comparing density in one and two dimensions because the 2D density values are equivalent to the 1D values plus a constant. As we see in Figure 5, the volume of speed in 2D is larger than 1D, because the magnitude of the speed in 2D is inherently greater than 1D. The significance of this difference can be further investigated, depending on quantitative analysis and the objective of the experiment (i.e., is the lateral displacement of the head important for the research?).

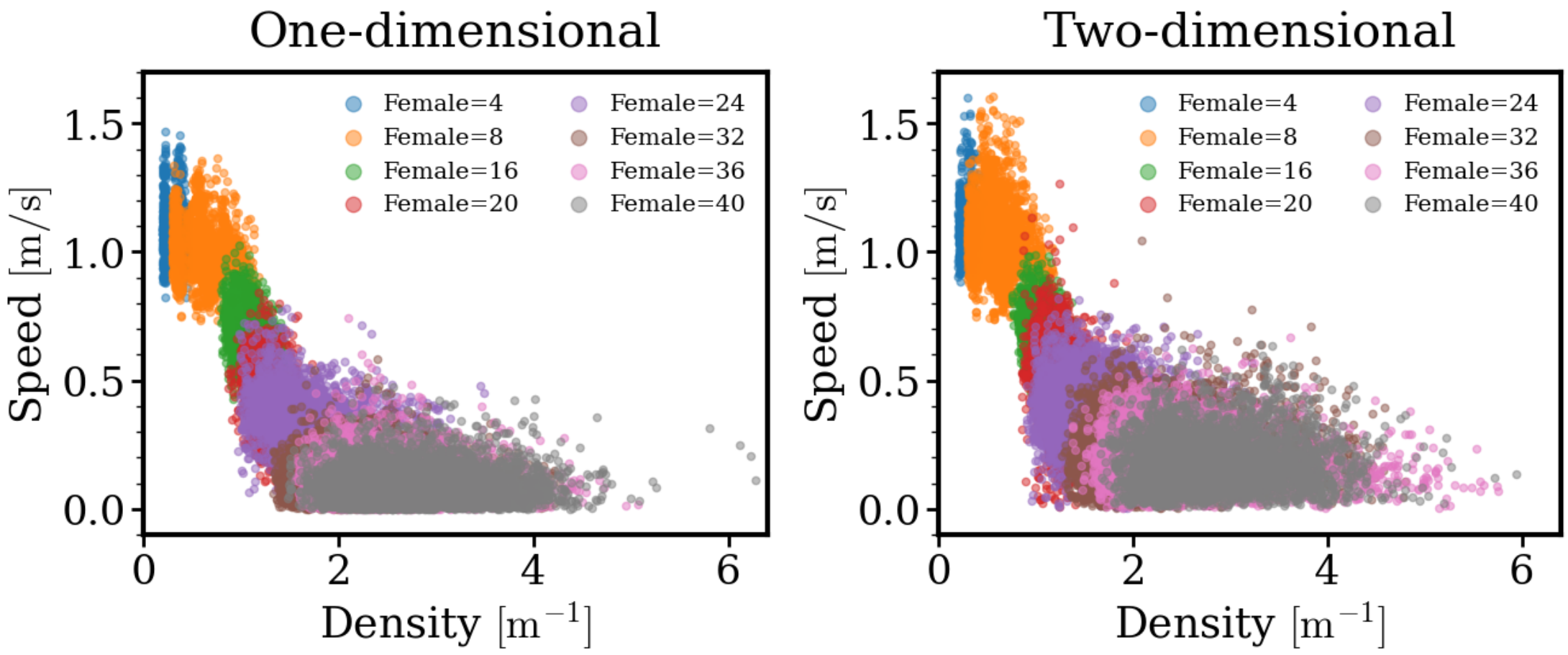

**Figure 5** Speed-density relation using 1D and 2D measurements of the speed.

The fourth aspect concerns the *phase of movement chosen for analysis*. In single-file experiments, Chattaraj et al. [2] identify three distinct phases of movement: *acceleration*, during which pedestrians start walking and their speed increases gradually; *steady state*, where their speed remains relatively stable; and *deceleration*, in which individuals gradually reduce their speed until they leave the setup or stop. Most studies focus on investigating movement characteristics during the steady state, except Wang et al. [30]. In the latter, the authors study how people slow down when walking in single-file to better understand their behavior during sudden stops [30]. The author's motivation for conducting this study is to enhance evacuation plans, prevent collisions, and ensure safety during emergencies. We believe that analyzing the data from a steady state allows gaining valuable insights into system behavior while simplifying the analysis. However, we assume it is essential to recognize the limitations of steady-state analysis and consider transient effects when necessary for a comprehensive understanding of pedestrian dynamic systems.

Finally, we summarize an artifact related to the calculation of movement quantities that influence single-file movement analysis as reported in the literature. Jelic et al. [10] demonstrate that the number of *detected markers* during data extraction affects the analysis. Some pedestrians' head markers are occluded in the experimental videos resulting in the loss of their head trajectories during specific time intervals. Jelic et al. compare the density-speed relationship using different numbers of detected markers and find that data points for all marker quantities mostly overlap. Additionally, density values are higher with fewer detected markers because density calculations include distances between pedestrians and their predecessors and followers. Hidden predecessors or followers not included in the trajectories increase these distances. We recommend that the position and numbers of detected markers match the real experiment's precision to avoid inaccuracies in the analysis.

## 6. Factors that influence movements

Various factors can be examined in pedestrians' single-file experiments (see Figure 7). In this section, we focus on discussing the influential factors already investigate on the ground-level evacuation scenario under closed-boundary conditions. We categorize these factors and discuss their influence on the characteristics of pedestrian movement.

Analyzing the impact of various influential factors is essential for modelers simulating pedestrian movement and for event organizers to implement safety procedures. To understand pedestrian walking behavior, we thoroughly explore potential factors and their impact on movement quantities, such as speed changes and flow variations. Analyzing these factors helps uncover correlations and causal relationships between variables, which are important for defining movement.

By observing experimental videos, participating in experiments, reviewing relevant literature, and conducting research on diverse aspects of single-file movement, we categorize these influential factors into three main groups based on their sources (see Figure 6):

- **Personal attributes** such as age, gender, etc.

- **Cognitive factors** involve mental processes and knowledge acquisition through thoughts, experience, and the senses, i.e., route choice, and motivation.

- **Social factors** including interactions with other pedestrians.
- **Environmental factors** including physical characteristics and layout of the experiment where individuals move and interact, such as location, weather, lighting conditions, etc.

We define *social conventions* as a set of agreed-upon or generally accepted standards and social norms that a group of people follows. These conventions influence walking behavior, as observed by Chattaraj et al. [2] in their pioneering research comparing young German and Indian participants. They conduct quantitative and qualitative analyses of the free-flow speed, density-speed, and speed-headway relations of Indian and German experiments. The results show that German walking speed is more dependent on density than Indian speed, with Indian data exhibiting greater fluctuations in speed and density (unordered behavior). Germans maintain greater personal space (headway distance) than Indians. Furthermore, both groups have similar free-flow speeds when walking alone. Bilintoh et al. [41] also examine social conventions by studying the *locations* of compatriots in single-file experiments conducted in Ghana and China with African students. They compare movement characteristics such as density, speed, flow, and headway. Their analysis reveals that Ghanaian pedestrians (speed between $0.74 \pm 0.01$ m/s and $0.32 \pm 0.02$ m/s) walk slower than the African students in China (speed between $1.11 \pm 0.01$ m/s and $0.31 \pm 0.03$ m/s) at the same global densities of $0.62\ \mathrm{m}^{-1}$ and $0.95\ \mathrm{m}^{-1}$, respectively. Additionally, Ghanaians maintain smaller personal space than African students in China based on headway distances.

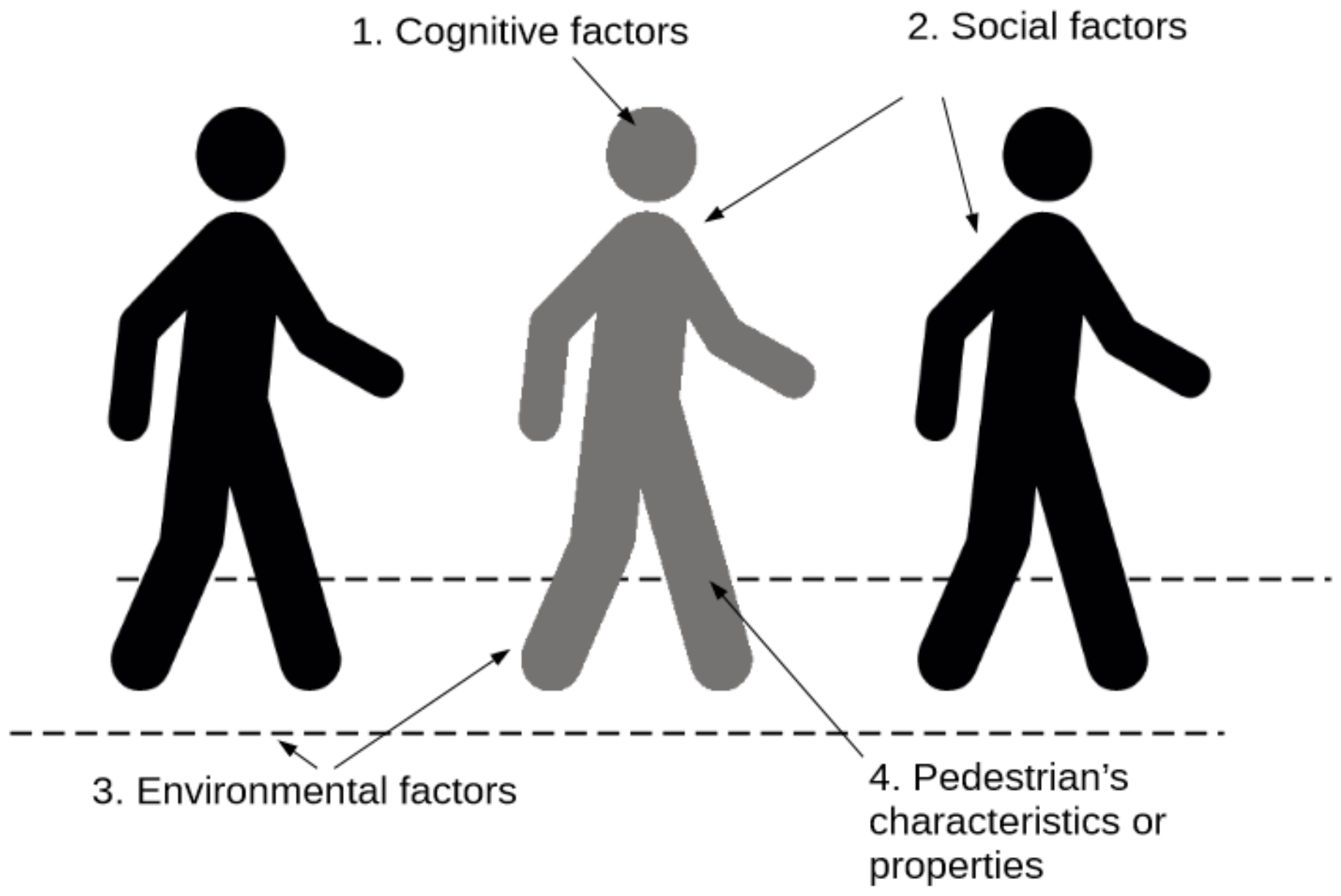

**Figure 6** Main groups of factors that Influence movement in single-file experiments as proposed by the authors

Age and gender are *personal attributes* influencing movement. Ren et al. [25] and Cao et al. [12] investigate the *age* effect on pedestrian dynamic. Cao et al. conduct a comparative analysis of homogeneous and heterogeneous age groups, including youth (16-18 years, average age 17), old adults (45-73 years, average age 52), and mixed groups (youth and elders randomly ordered). In contrast, Ren et al. focus on elders aged 50-85 years, with an average age of 70. Cao et al. find that young students move faster than old adults in the speed-density relationship. At the same density, the young group is faster than the mixed group. The mixed group's speed is slightly lower than that of the old adults' group at densities between 0.5 $m^{-1}$ and 1.2 $m^{-1}$, while it is higher at densities below 0.5 $m^{-1}$. Additionally, flow increases monotonically with density for all groups but reaches different peak flows: 1.3 $s^{-1}$ for youth, 0.9 $s^{-1}$ for old adults, and 0.7 $s^{-1}$ for mixed groups around a density of 0.9 $m^{-1}$. Ren et al. compare the speeds of elders and old adults, finding that the elders walk slower than the adults in the low-density scenarios but at roughly the same speed in the mixed group. Furthermore, stop-and-go waves occur frequently and last for a longer duration in the elderly group compared to the old adult group. The authors observe from the experiment videos, time-space diagrams, and headway values that some elders wait several seconds until they have a certain sufficient distance in front to move again. Elders do not resume walking synchronously with the preceding pedestrian after stopping, a phenomenon they term "active cease". We attribute these differences in movement to the physical mobility capabilities of pedestrians.

In *gender* studies, Subaih et al. [56] and Paetzke et al. [39] explore the impact of gender composition on pedestrian movement. While their objectives are similar, their contributions and findings differ. Both studies use statistical analyses to assess the significance of their findings with different testing methods. They find that homogeneous gender groups (either all female or all male) exhibit similar density-speed relationships. Subaih et al. observe differences in the density-speed diagram between homogeneous and heterogeneous (mixed-ordered) gender groups, suggesting that the gender of neighboring pedestrians affects movement. In contrast, Paetzke et al. expand the analysis to various group compositions and find that gender composition effects on speed-density relations are either nonexistent or only present within a narrow density range. They attribute these discrepancies to different statistical methods and data preparation. Furthermore, Paetzke et al. investigate additional factors like *weight*, *height*, and the gender of the preceding pedestrian but conclude that these factors do not significantly improve the predictability of pedestrian speed. This reinforces that gender composition and these additional factors have minimal impact on pedestrian dynamics in single-file movement.

Some influential factors are controlled or manipulated to observe their effects on the experimental results (*motivation*). For example, organizers use instructions, music, and environmental changes to assess their impact on participants' behavior and walking patterns. Lu et al. [33] investigate pedestrian movement under different *social distancing measures* similar to those during COVID-19 in China: 1 m, 2 m, and normal conditions (before COVID-19). They find that social distancing measures caused participants to maintain greater distances than normal conditions, though some violations occurred. Stop-and-go waves under social distancing measures are observed not only at high densities but also at low-density ranges. We suppose the reason is that pedestrians prefer to stay alert and maintain the predefined distance to follow the instructions. Thus, they stop to estimate and adjust the distance headway before proceeding. Wang et al. [30] investigate the effect of *stop distances* by instructing participants to either stop close to or normally behind their predecessors. The close-stop condition results in shorter average stop distances (0.34 m) compared to normal stops (0.63 m). Additionally, the speed-distance headway slope is steeper in close-stop experiments, indicating more abrupt deceleration as participants approach the person in front.

Appert-Rolland et al. [22] study the *cognitive processes* of pedestrians, focusing on how increased freedom of movement affects pattern formation, interaction, and decision-making in crowds using a single-file system. In their experiments, participants are instructed to walk in a self-chosen virtual circle without predefined boundaries. Consequently, participants form circular paths by following and interacting with their predecessors (following behavior).

Another group of researchers focuses on the influence of *music, songs, and metronome* rhythm on pedestrian motion. They hypothesize that music and rhythm enhance pedestrian flow in congested situations without causing danger. Zeng et al. [28] perform an oval experiment to understand the impact of background music on movement. Seven experiments are performed: three with different music tempos, three with rhythms from a metronome device, and one without music (normal conditions). The authors only analyze and compare the movement under normal conditions and with music at 120 beats per minute (BPM). The analysis of density-speed and density-flow shows that at the medium and high densities investigated, speed and flow are lower with background music than under normal conditions. Stop-and-go waves appear in both cases at a global density of $\rho_{\text{glob}} = 1.82\ \text{m}^{-1}$, but with background music participants stop frequently and for a longer duration.

In studying the impact of metronome rhythm, Yanagisawa et al. [9], Ikeda et al. [16], and Li et al. [40] conduct experiments with different types of setups with and without a rhythm of 70 BPM. Yanagisawa et al. use experimental data to validate their pedestrian flow model, which combines two primary parameters: step size and walking pace (steps per unit time). Their results indicate that the slower walking rhythm can enhance pedestrian flow in congested environments. This improvement occurs because pedestrians maintain a more consistent pace and avoid abrupt reductions in step size, which is observed in the experimental data. Specifically, the slow rhythm helps synchronize pedestrian movement, reducing variability and improving flow at high densities.

Ikeda et al. analyze the impact of steady beats on the cognitive processes of pedestrians by measuring participants' frontopolar brain activity in walking and stepping groups. They find that playing a steady beat sound (like a metronome) helps groups walk together more smoothly in crowded situations and improves the coordination between their brain activities, particularly in the prefrontal region. The aforementioned research on music and metronome rhythms demonstrates that pedestrian flow can be improved by music and rhythm, which influence stepping behavior and cognitive processes.

There are also *Environmental factors* that significantly impact pedestrian movement, as demonstrated by various studies. Cao et al. [23] investigate the movement under various *visibility conditions* by testing three levels of light transmission (0.3%, 0.1%, and 0.0%). The study show that pedestrian speed and flow change significantly with different visi-bility conditions. Specifically, the following behavior (toward proceeding pedestrians or the walls) is observed at light transmissions of 0.1% and 0.0%. Additionally, stop-and-go waves appear at low densities and increased as visibility decreased. The maximum specific flow rates vary with visibility, being 1.3 $s^{-1}$, 1.15 $s^{-1}$, and 0.9 $s^{-1}$ for light transmissions of 0.3%, 0.1%, and 0.0%, respectively.

Chattaraj et al. [2] investigate the influence of *corridor length* and found no significant impact on speed-density or speed-headway distance relations. Jelic et al. [10] analyze how the *walking path* -either along the inner wall or the outer wall of a circular setup-affects pedestrian movement. The authors observe that pedestrians maintain a slightly greater distance from the wall when walking along the outer path compared to the inner path. Furthermore, they find no significant differences in density-speed relations between the two paths.

Ren et al. [25] explore the effects of *vertical walls* in various experimental setups, observing different pedestrian behaviors based on wall presence. They examine three cases: case one with a wall on one side of a straight section, case two with walls on both sides of a straight section, and case three with no walls in curved sections. Pedestrians in case one tend to walk away from the wall towards the open side, while movements in case two are less fluctuating and more concentrated compared to cases one and three. In case three, fluctuations are more frequent, and pedestrians often crossed boundaries, especially at high densities, leading to overlapping. The study concludes that boundary types, whether vertical walls or ground tape, significantly affect pedestrian movement characteristics. Ren et al. [25] also observe the influence of the *setup shape* on the speed of pedestrians (straight and curved). This influence is further analyzed by Fu et al. [32], where the authors find that pedestrian flow increases in the straight part (oval experiments) than the flow in the curve part (discussed before in Section 3).

Ma et al. [29] conduct experiments to understand the impact of *height constraints* (1.0 m, 1.2 m, 1.4 m, 1.6 m, and 2.0 m) on pedestrian movement. The authors find that speed distributions across different heights follow a Gaussian pattern, with lower height constraints significantly reducing pedestrian speeds and altering the flow. In conclusion, experimental settings such as visibility, corridor length, walking path, boundaries, setup shape, and height constraints significantly affect pedestrian movement analysis. These factors should be carefully considered in the analysis and interpretation of results.

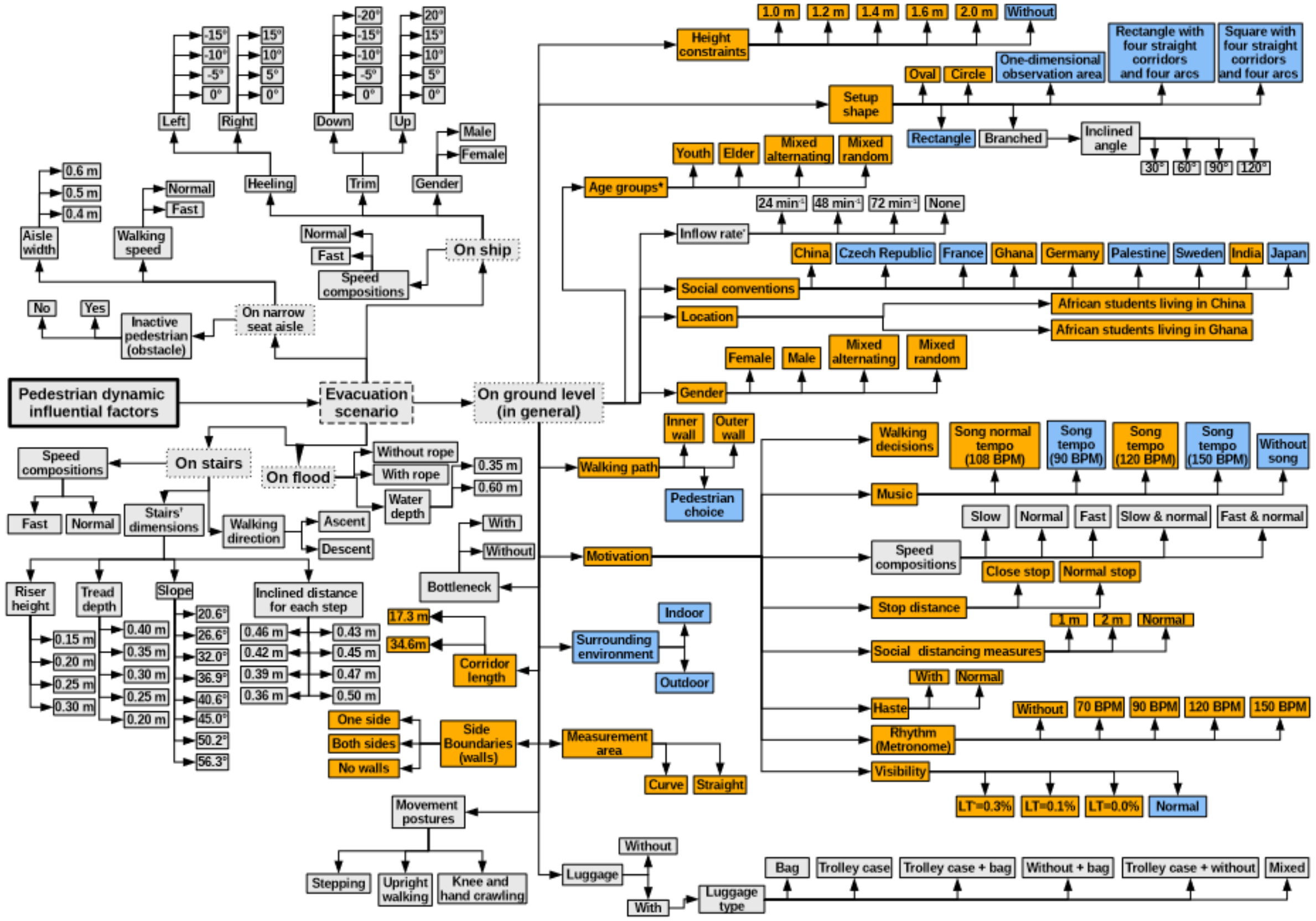

**Figure 7** Categories of factors influencing the movement in single-file experiments from the literature. Arrows indicate the categories of influential factors concerning the evacuation scenario, and the influential factors along with lists of quantitative or categorical variables for comparison. Orange denotes factors studied under closed boundary conditions at ground level, while blue represents factors assumed to affect the movement (by the authors).

# 7. Methodology for Preparing Trajectory Data and Calculating Movement Quantities

From reviewing the literature, we notice that various studies employ different codes and tools to analyze experiments. These differences stem from diverse experimental setups and settings. We analyze data from multiple studies to ensure a comprehensive understanding, i.e., comparing the one-dimensional and two-dimensional measurements discussed in Section 5. We identify the need for foundational software for single-file experiments that researchers can build upon. This software is open-source and available online, enabling developers to systematically analyze experiments across different settings. The tool serves as a standardized approach for data analysis. Furthermore, it provides a foundation for future development, allowing other researchers to enhance its capabilities by adding new features.

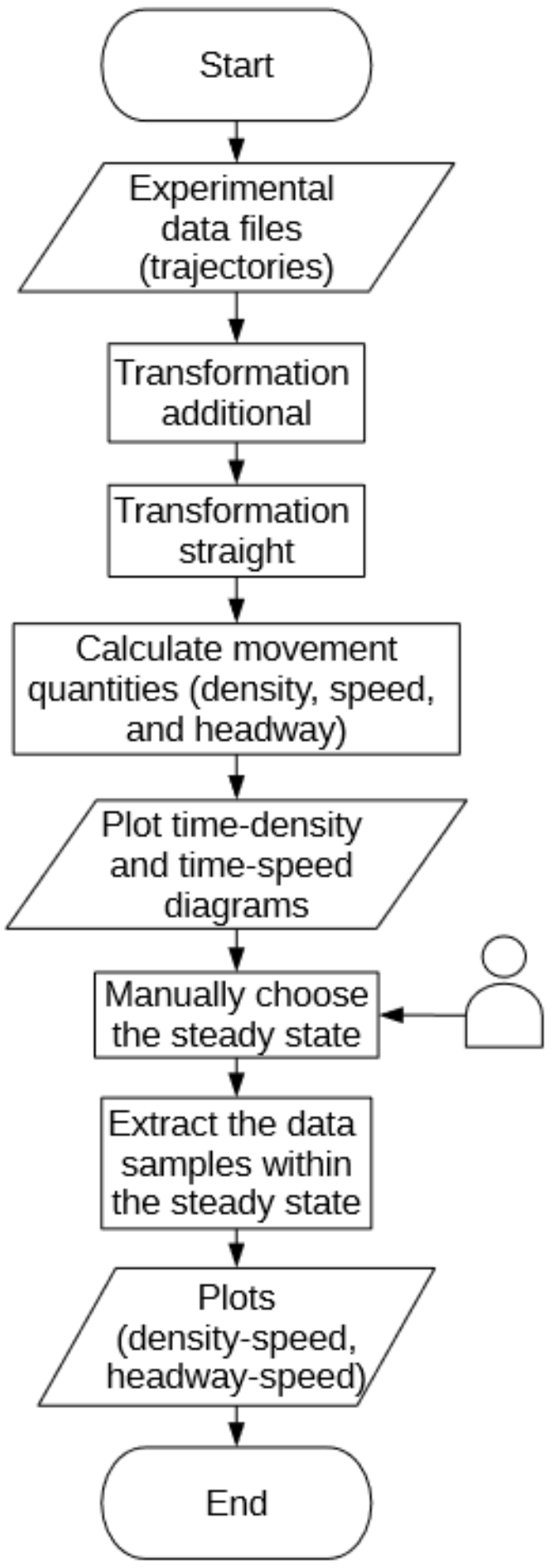

**Figure 8** Flowchart for calculating movement quantities using head trajectories.

In this section, we introduce a Python tool for analyzing single-file experiments. We also propose a methodology for preparing experimental data (head trajectories), calculating movement quantities, and analyzing the common relations investigated in the single-file literature: density-speed and density-headway. To qualitatively and quantitatively analyze the single-file movement by using head trajectories, we propose the methodology outlined in the flowchart presented in Figure 8 to prepare the raw data and calculate movement quantities.

The first two steps in the methodology for preparing the raw trajectory data are *transformation additional* and *transformation straight*. Upon observing the plots of raw trajectories in the literature, we notice that the $(x,y)$ values are centered around different points, depending on the trajectory extraction process (location of the coordination system). To convert oval trajectories into straight - a process we refer to as the "transformation straight" step, following the method of Ziemer et al. [13] - we adjust the trajectories to a new, unified Cartesian coordinate system, $T : \mathbb{R}^2 \to \mathbb{R}^2, \begin{pmatrix} x \\ y \end{pmatrix} \mapsto \begin{pmatrix} x' \\ y' \end{pmatrix}$.

In this system, trajectories represent a person starting her/his walk from the beginning of the bottom straight corridor $(x = 0)$, along the corridor's central line $(y = 0)$ (Sub-figure 9(b) show the new coordination system). *Additional transformation* is achieved by applying appropriate transformations in geometry, such as rotation, shifting, etc (see Sub-figures 9(a) transform to 9(c)).

Some common cases for additional transformation are summarized as follows:

1. In some experiments, the $(x,y)$ coordinates are given in centimeters. We convert them to meters by setting the unit conversion factor $u$ as follows: if the original units are in centimeters, then $u = 100$ to convert to meters; otherwise, $u = 1$.

2. To ensure the straight segments of the oval setup are parallel to the x-axis, rotate the trajectories by 90° clockwise, transforming $(x,y) \to (y,-x)$, or 90° anticlockwise, transforming $(x,y) \to (-y,x)$. For experiments, pedestrians walk either clockwise or anticlockwise. In clockwise experiments, apply horizontal reflection to calculate distances, setting constraints $i = -1$ and $j = -1$ for axis reflections; otherwise, set $i = 1$ and $j = 1$.

3. To align the origin with the middle line of the corridor, as shown in Sub-figure 9(b), we need to shift the trajectories horizontally or vertically. For horizontal and vertical translations, we use the constants $k \in \mathbb{R}$ and $d \in \mathbb{R}$, respectively.

The additional transformation equations $T$ are:

$$x' = \frac{i.x}{u} + k. \tag{1}$$

$$y' = \frac{j.y}{u} + d. \tag{2}$$

In case we want to calculate the movement quantities for pedestrians walking inside a specific area (the straight part), we need to extract $(x,y)$ values from within the space interval of the measurement area, $(x,y) \in [a,b]$, where $a \in \mathbb{R}$ and $b \in \mathbb{R}$ represent the minimum and maximum x-axis values, respectively.

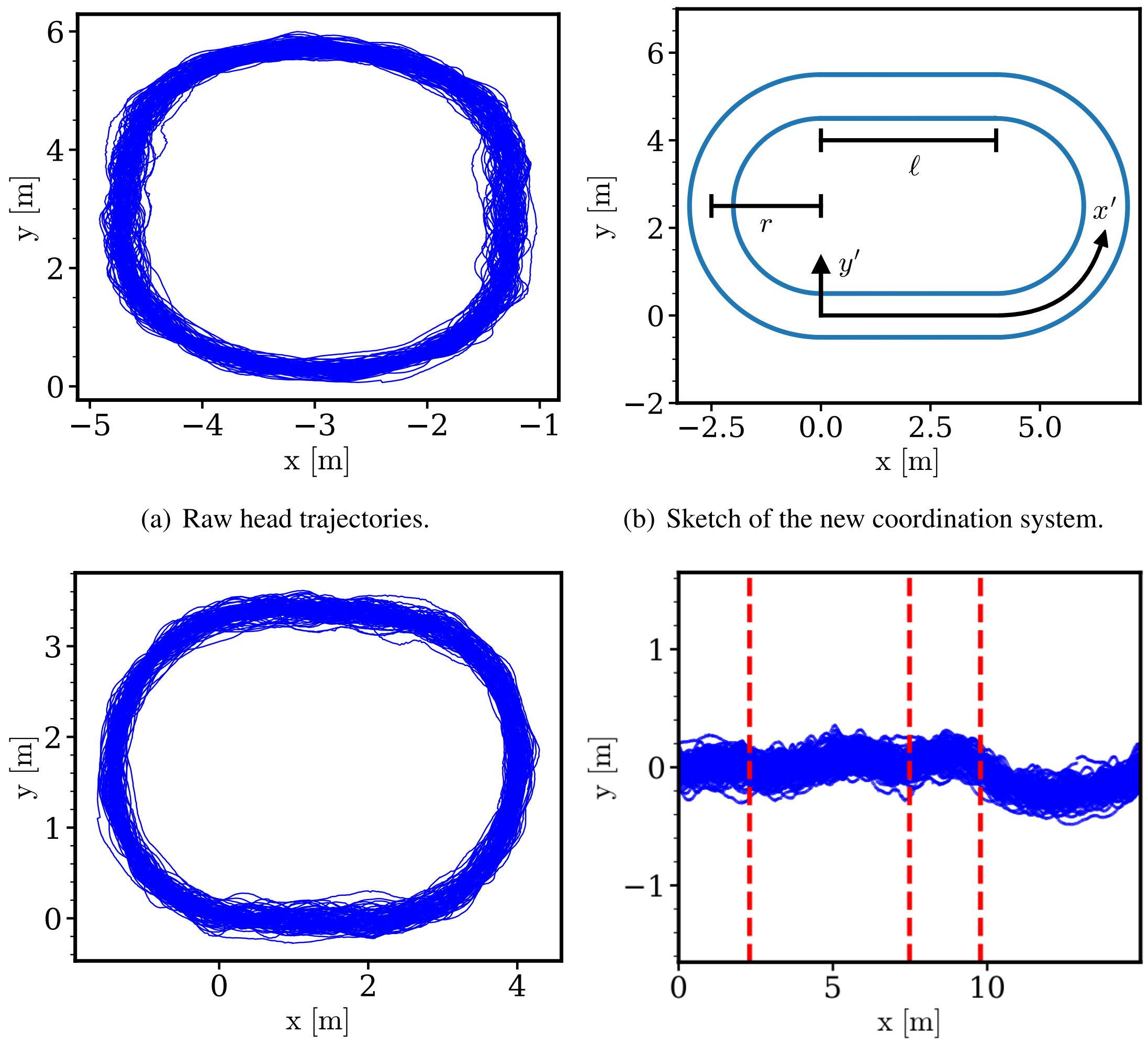

(a) Raw head trajectories.

(b) Sketch of the new coordination system.

(c) Transformation additional.

(d) Transformation straight.

**Figure 9** The steps of the transformation applied to the trajectory data extracted from the single-file experiment.

To apply *transformation straight*, $T' : \mathbb{R}^2 \to \mathbb{R}^2$, $\begin{pmatrix} x' \\ y' \end{pmatrix} \mapsto \begin{pmatrix} x'' \\ y'' \end{pmatrix}$, the equations are defined as follows:

$$x'' = \begin{cases} \ell + r \text{ arcos}\left(\frac{r-y}{\sqrt{(x-\ell)^2+(y-r)^2}}\right) & x > \ell, \\ x & 0 \leq x \leq \ell,\ y < r, \\ 2\ell + r\pi - x & 0 \leq x \leq \ell,\ y \geq r, \\ 2\ell + r\pi + r \text{ arcos}\left(\frac{-r+y}{\sqrt{x^2+(y-r)^2}}\right) & x < 0, \end{cases} \tag{3}$$

and

$$y'' = \begin{cases} \sqrt{(y-r)^2} - r & 0 \leq x \leq \ell, \\ \sqrt{(x-\ell)^2 + (y-r)^2} - r & x > \ell, \\ \sqrt{x^2 + (y-r)^2} - r & x < 0, \end{cases} \tag{4}$$

where $\ell$ represents the length of the straight segment of the oval corridor, and $r$ denotes the radius of the curved part (see Figure 9(b)).

The third step of the methodology involves calculating the movement quantities. In our paper, we calculate *Voronoi 1D density*, *individual instantaneous speed*, and *headway distance*. To calculate the *headway distance*, we apply the following equation:

$$h_i(x) = \begin{cases} x_{i+1}(t) - x_i(t) & i = 1, \ldots, n-1, \\ (c - x_n(t)) + x_1(t) & i = n. \end{cases} \tag{5}$$

where $n$ is the number of pedestrians in the experiment and $c$ is the geometry circumference.

*Voronoi 1D density* is defined as:

$$\rho_i(x) = \begin{cases} \frac{2}{h_{i-1}(x)+h_i(x)} & x \in [\frac{h_{i-1}(x)}{2}, \frac{h_i(x)}{2}[, \\ 0 & \text{Otherwise}. \end{cases} \tag{6}$$

Finally, we calculate the *individual instantaneous speeds* of pedestrians using the following equation for *side-view experiments* or analysis within a specific measurement area:

$$v_i(t) = \begin{cases} \frac{x_i(t+\Delta t/2) - x_i(t-\Delta t/2)}{\Delta t} & t + \Delta t/2 \leq t_{\text{end}},\ t - \Delta t/2 \geq t_{\text{start}}, \\ \frac{x_i(t_{\text{start}}) - x_i(t-\Delta t/2)}{t - t_{\text{start}} + \Delta t/2} & t + \Delta t/2 > t_{\text{end}},\ t - \Delta t/2 \geq t_{\text{start}}, \\ \frac{x_i(t+\Delta t/2) - x_i(t_{\text{end}})}{t_{\text{end}} - t + \Delta t/2} & t + \Delta t/2 \leq t_{\text{end}},\ t - \Delta t/2 < t_{\text{start}}, \\ 0 & \text{otherwise}, \end{cases} \tag{7}$$

where $t_{\text{start}}$ and $t_{\text{end}}$ are the time when the pedestrian $i$ enters and leaves the measurement area, respectively. The short time constant of $\Delta t = 0.4$ s (10 frames) is used to smooth trajectories and avoid fluctuations in the stepping behavior of pedestrians. For the *top-view experiments*, Equation 7 (case one) is used to calculate the *1D* individual instantaneous speeds. In *2D*, the speed is calculated by dividing the displacement in 2D by $\Delta t$ as follows:

$$v_i(t) = \frac{\sqrt{(x_i(t+\Delta t/2) - x_i(t-\Delta t/2))^2 + (y_i(t+\Delta t/2) - y_i(t-\Delta t/2))^2}}{\Delta t}. \tag{8}$$

For more details regarding the proposed analysis tool, check the GitHub project *SingleFileMovementAnalysis* [4]. The tool is tested across 10 experiments involving 28 datasets, as detailed in Appendix C, Table 5.

## 8. Summary, trends and future outlooks

In this section, we highlight the trends and the directions of future research in single-file experiments.

In single-file systems (discussed in Section 2), cars and bicycles are influenced by mechanical effects and inertia. This leads to systematic speed control and the maintenance of space to prevent collisions, especially in traffic jams. In contrast, animals such as mice and ants do not maintain personal space or time gaps and often overlap, pause randomly, and even move backward, resulting in higher flow at crowded densities. Pedestrians, like vehicles, maintain personal space, but like animals, their movements are flexible. At high densities, people slow down to avoid contact, creating stop-and-go patterns driven by psychological and physiological factors. These differences should be confirmed by further experiments.

In analyzing pedestrian movement, there is a clear consensus regarding the choice of experimental setups as reviewed in Section 3. Closed systems are preferred for investigating the fundamental diagram, as they provide controlled conditions that eliminate external boundary effects and control the density. In contrast, open systems are valuable for examining external influences, such as pedestrians' streams from different directions or decision-making in movement (i.e., direction to take, following behavior), which are critical in real-world scenarios. However, the influence of boundaries in these setups remains an area that requires further investigation. Understanding how boundary conditions affect pedestrian dynamics can enhance our comprehension of movement.

Data collection in pedestrian studies is develop to become more comprehensive, incorporating surveys that gather meta-data about pedestrians, such as demographic details and socio-psychological contexts. This trend reflects a growing recognition of the importance of understanding not only the physical movement of individuals but also the factors influencing their behavior. Furthermore, automation through tools like PeTrack has significantly advanced the precision of head trajectory extraction from videos, allowing for a more detailed analysis of pedestrian movement. Different types of data are extracted by Thompson et al. [38] such as the trajectories of shoulders, hips, knees, tips of the toes, and heels to improve the calculation of stepping quantities. By integrating this additional information, researchers can gain deeper insights into pedestrian dynamics and the various influences that shape movement in different environments. A new technologies to capture, track and analyze pedestrian steps, such as step extent and step frequency is required. In the literature, the measurements of stepping behavior are mostly analyzed using the head trajectories. This will enable researchers to improve accuracy and reduce errors in the analysis.

In Section 6, we discuss several factors that influence pedestrian dynamics which have already been investigated in the literature, categorizing them into personal attributes (age, gender), cognitive factors (route choice and motivation), social factors (interactions with others), and environmental factors (visibility and layout). Each of these factors significantly impacts, or does not, movement characteristics such as speed and flow, highlighting the complexity of pedestrian behavior. There is a growing interest in research examining the effects of rhythm and music on movement to improve the flow of pedestrians in congested situations. The researches [9, 16, 28, 40] highlight how the rhythm and music influence the fundamental diagram and the pedestrian behavior. Investigating how varying tempos affect coordination and flow in different environments may provide valuable insights into optimizing pedestrian movement, particularly in crowded settings where synchronization can enhance safety and efficiency. Future research could also explore the effect of the surrounding environment (indoor or outdoor) on pedestrian dynamics. Additionally, investigating the impact of emerging technologies, such as wearable devices that provide real-time data on movement, could further enhance our understanding of pedestrian dynamics. Finally, focusing on factors such as the role of individual cognitive processes and socio-psychological factors in pedestrian behavior. For example, how individuals feel under different motivations (e.g., evacuation, noise, rhythm) and how these factors influence their decision-making. More suggested factors are highlighted in Figure 7.

## 9. Conclusion

This article comprehensively reviews the literature on single-file pedestrian movement, with a focus on experiments and data analysis. We provide a scientific background and discuss the significance of single-file experiments in pedestrian dynamics. Then, we compare different traffic systems - including humans, mice, rats, bicycles, and cars - to highlight their similarities and differences. From this comparison, we derive insights that contribute to our understanding of pedestrian dynamics. Furthermore, we present a detailed discussion and categorization of the types of experimental setups, data collection methods, movement quantities, and influential factors of the movement, and provide our discussion. Finally, we propose a methodology and introduce the "SingleFileMovementAnalysis" tool for analyzing single-file pedestrian dynamics. After the comprehensive review, we recognize the ongoing need for further research in single-file movement. Specifically, experimental research focuses on the cognition processes of moving pedestrians to understand the related factors influencing the dynamics. We also suggest performing research concerning defining and automating the steady-state in pedestrian single-file movement. Moreover, we encourage further experiments to investigate new influential factors and validate new data collection devices. There is still room for improvement in research on pedestrian single-file movement to compare experimental data against existing research objectively and easily, thereby improving the quality of analysis.

**Acknowledgements** The authors would like to thank Yi Ma, Guang Zeng, Jun Zhang, and Jian Ma for sharing their datasets of single-file experiments. Furthermore, we thank Tobias Schrödter for valuable feedback regarding the Python analysis tool. This work is supported by the German Federal Ministry of Education and Research (BMBF: funding number 01DH16027) within the framework of the Palestinian-German Science Bridge project. The authors acknowledge the Franco-German research project MADRAS founded in France by the Agence Nationale de la Recherche (ANR, French National Research Agency), grant number ANR-20-CE92-0033, and in Germany by the Deutsche Forschungsgemeinschaft (DFG, German Research Foundation), grant

number 446168800.

**Author Contributions** Rudina Subaih: Conceptualization, Methodology, Software, Validation, Formal analysis, Investigation, Data curation, Visualization, Writing –original draft, Writing – review & editing. Antoine Tordeux: Conceptualization, Writing – review & editing, Supervision. Mohcine Chraibi: Conceptualization, Writing – review & editing, Supervision.

# A. Single-file Experiments

**Table 3** Overview of the experimental publications reviewed by the authors that focus on single-file experiments for various traffic systems.

| | Experiment (AuthorYear [cite]) | Snapshot | Objects | Country / territory (where the experiment performed) | Environment | Setup shape / type | Main contribution * | Main result * |
|---|---|---|---|---|---|---|---|---|
| 1 | Seyfried2005 [1] | 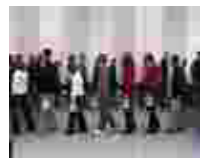 | Pedestrians | Germany | Indoor | Oval | Investigate the velocity-density relation of unidirectional movement for pedestrians in different global densities | A linear relationship between the average velocities and the headway distances |
| 2 | Sugiyama2008 [48] | 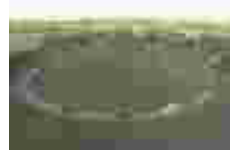 | Cars (human-driven) | Japan | Outdoor | Circle | Experiment on the emergence of traffic jams in vehicle dynamics | Traffic jams emerge in a single-file system when vehicles move in a circle |
| 3 | Chattaraj2009 [2] | 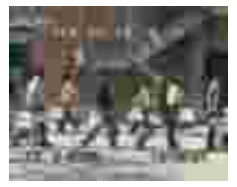 | Pedestrians | India | Outdoor | Oval | Examine the impact of cultural factors and corridor length on the fundamental diagrams | The cultural factor influences the fundamental diagram, whereas the corridor length has no impact |
| 4 | Lukowski2009 [5] | 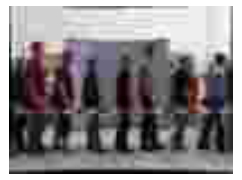 | Pedestrians | Germany | Indoor | Oval | Investigation of the influence of motivation (normal walking and walking under urgency) on pedestrian movement | Motivation affects pedestrian movement with different conditions leading to distinct fundamental diagrams |
| 5 | Liu2009 [6] | 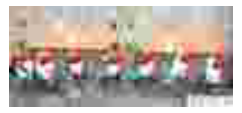 | Pedestrians | China | Outdoor | Oval | Present and examine a pedestrian detection and tracking tool, along with a quantitative analysis of the microscopic characteristics of pedestrians | An automated tool using the mean-shift algorithm to extract head trajectories and quantify microscopic characteristics of pedestrians such as velocity, density, and lateral oscillation |
| 6 | Jezbera2010 [7] | 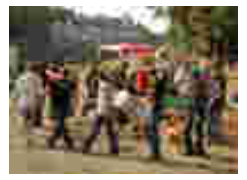 | Pedestrians | Czech Republic | Outdoor | Circle | Describe the statistical properties of headway distances using the Gaussian Unitary Ensemble of random matrices | Description of headway as derived from mathematical models of one-dimensional random walkers |
| 7 | Seyfried2010 [8, 58] | 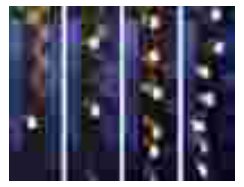 | Pedestrians | Germany | Indoor | Oval | Analysis of pedestrian movement in congested situations | Coexistence of moving and stopping states as demonstrated by the velocity-density relationship |
| 8 | Yanagisawa2012 [9] | 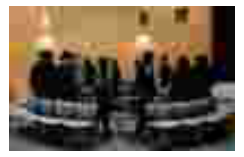 | Pedestrians | Japan | Indoor | Circle | Propose a velocity model incorporating step size and walking pace | Rhythm improves flow in congested situations |
| 9 | Jelic2012 [10, 54] | 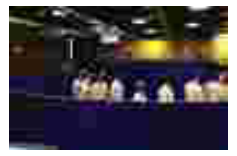 | Pedestrians | France | Indoor | Circle | Focus on the fundamental diagram (relationship between instantaneous velocity and headway distance) | The speed-headway exhibits a piecewise linear behavior in the large densities |
| 10 | Song2013 [11] | 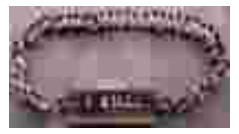 | Pedestrians | China | Outdoor | Oval | Improve a one-dimensional continuous distance model by introducing desired direction and analyzing microscopic movement characteristics | The proposed model describes the one and two-dimensional movement |
| 11 | Tadaki2013 [49] | 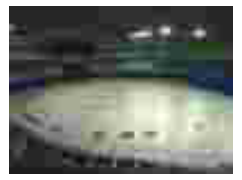 | Cars (human-driven) | Japan | Indoor | Circle | Investigate how traffic jams can emerge even without bottlenecks at a certain high density | Jams occur at high densities, while free flow is maintained at low densities |
| 12 | Zhang2014 [46] | 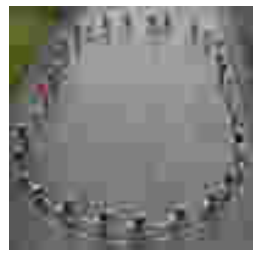 | Bicycles | Germany | Outdoor | Oval | Identify a universal flow-density relationship for bicycle, pedestrian, and car movement systems | Despite different behaviors, the flow-density relationship is described by a universal fundamental diagram after rescaling |
| 13 | Cao2016 [12] | 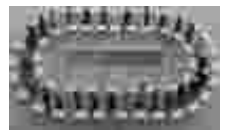 | Pedestrians | China | Outdoor | Oval | Investigate movement characteristics across different age groups | Age composition influences velocity and the occurrence of stop-go waves |
| 14 | Ziemer2016 [13] | 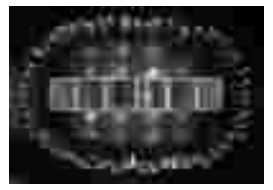 | Pedestrians | Germany | Indoor | Oval | Analyze congestion dynamics using complete trajectory data (both straight and curved sections) | Describe the stop-and-go waves and fundamental diagrams, which have the same shape in both straight and curved sections of the setups |
| 15 | Zhao2017 [14] | 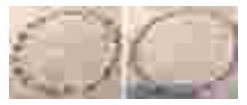 | Pedestrians, Bicycles | China | Outdoor | Circle | Compare and study the space-time trajectories and space-velocity diagrams of pedestrians, bicycles, and vehicle | Similarities exist between the space-time trajectories and space-velocity diagrams of traffic systems |
| 16 | Jiang2017 [47] | 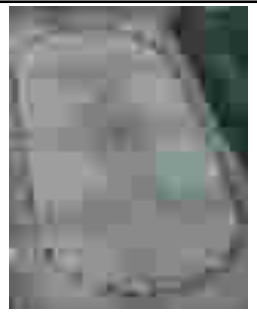 | Bicycles | China | Outdoor | Oval | Study the characteristics of bicycle movement | Present the bicycle numbers-flow relation and analyze the spatiotemporal evolution of bicycle flow |
| 17 | Chen2017 [15] | 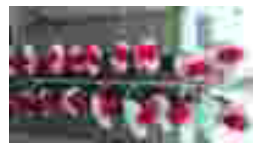 | Pedestrians | China | Indoor | Stairs | Investigate movement characteristics when ascending and descending stairs | Similar properties and different resolutions for density-speed and speed-headway diagrams are observed in ascending and descending movement |
| 18 | Ikeda2017 [16] | 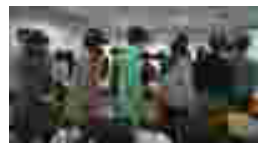 | Pedestrians | Japan | Indoor | Circle | Show that a steady beat can enhance coordinated group walking and inter-subject neural synchrony in congested situations | A steady beat significantly improved walking flow and increased neural synchrony between individuals |
| 19 | Gulhare2018 [17] | 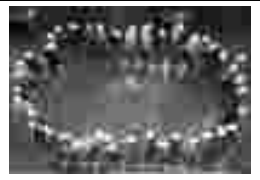 | Pedestrians | India | Outdoor | Oval | Compare data from a field study with controlled experiment | The data from experiments are different from the field data |
| 20 | Huang2018 [18] | 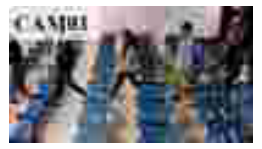 | Pedestrians | China | Indoor | Seat aisle | Investigate movement in narrow seat aisle | The speed of pedestrians changes with variations in aisle width |
| 21 | Ma2018 [19] | 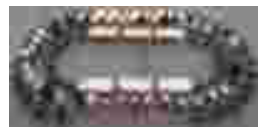 | Pedestrians | China | Outdoor | Oval | Analysis of continuous stepping behaviors in interacting pedestrians | The relationship between step length and headway distance shows a piecewise linear behavior |
| 22 | Sun2018 [20] | 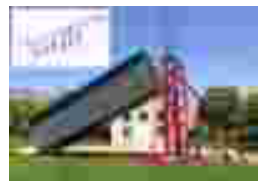 | Pedestrians | China | Indoor | Ship corridor | Investigate the effect of ship trim and heeling in movement characteristics | Trim angles impact pedestrians' speed more than heeling angles, which show less influence |
| 23 | Wang2018 [21] | 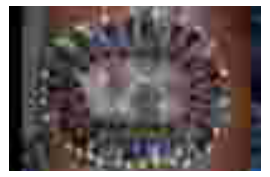 | Pedestrians | Germany | Indoor | Oval | Investigate the link between stepping and flow characteristics | There is a dependence between stepping locomotion and speed |
| 24 | Stern2018 [50] | 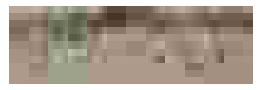 | Cars (autonomous vehicle) | USA | Outdoor | Circle | Reduce stop-and-go waves with intelligent control of autonomous vehicles | Reduce stop-and-go waves by controlling the velocity of a single vehicle in the system |
| 25 | Appert-Rolland2018 [22] | 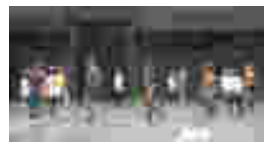 | Pedestrians | France | Indoor | Circle (virtual, without predefined path), one-dimensional observation area | Examine how pedestrians adapt their trajectories in crowds | Pedestrians adjust their headway according to the leader's speed |
| 26 | Cao2019 [23] | 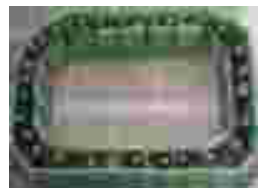 | Pedestrians | China | Outdoor | Rectangle with four straight corridors and four arcs | Understand the movement characteristics under limited visibility (light conditions) | Visibility conditions affect the occurrence of stop-and-go waves and the movement properties |
| 27 | Jin2019 [24, 55] | 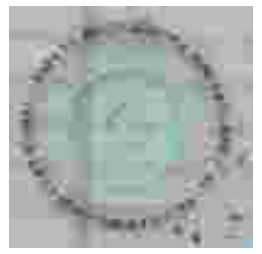 | Pedestrian | China | Outdoor | Circle | Investigate movement characteristics in a high-density circular setup (4 pedestrians per square meter) | Quantitatively discuss the values of density and speed for the stop and moving states, as well as the critical density values |
| 28 | Ren2019 [25] | 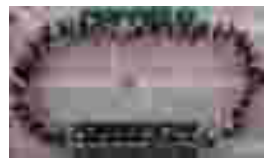 | Pedestrian | China | Outdoor | Oval | Compare the movement characteristics of different age groups, with a focus on the elderly | The elderly stop for longer periods and walk more slowly than other age groups |
| 29 | Subaih2019 [26, 56] | 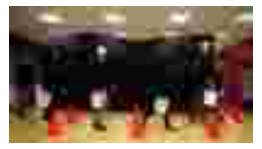 | Pedestrians | Palestine | Indoor | Oval | Investigate the influence of gender composition (male, female, mixed-random) on movement | Male and female groups walk at similar speeds, but they adjust their speed when moving together |
| 30 | Xiao2019 [43] | 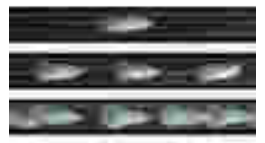 | Mices | China | Indoor | One-dimensional observation area | Investigate the movement characteristics of mice | A decaying velocity-density relationship at low densities and a non-strict relationship at high densities |
| 31 | Huang2019 [27] | 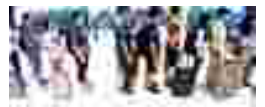 | Pedestrians | China | Indoor | One-dimensional observation area | Analysis of movement characteristics for pedestrians carrying luggage | Luggage has a minor influence on individual speed but a significant impact on headway distances |
| 32 | Zeng2019 [28, 60, 61] | 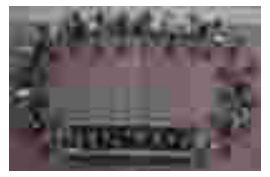 | Pedestrians | China | Outdoor | Oval | Compare the movement characteristics with and without background music | With music there is a higher frequency of stop-and-go waves, longer headway distance and stopping durations |
| 33 | Ma2020 [29] | 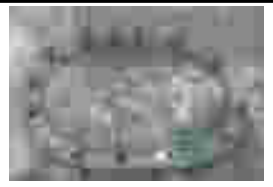 | Pedestrians | China | Outdoor | Oval | Investigate movement characteristics under different height constraints | Height constraints affect movement characteristics |
| 34 | Wang2020 [44] | | Ants | China | Indoor | One-dimensional observation area | Investigate ant movement under high-temperature stress | Random pauses and overtaking are observed among the ants |
| 35 | Wang2020a [30] | 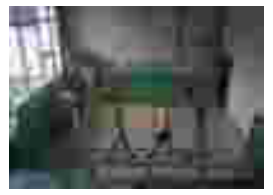 | Pedestrians | China | Indoor | Rectangle | Investigate the impact of different stopping distances (normal, close) on movement | The type of stop influences the speed |
| 36 | Wang2020b [31] | 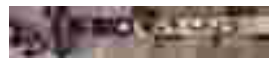 | Pedestrians | China | Outdoor | One-dimensional observation area | Investigate knee and hand crawling behavior during building evacuations in a fire scenario for different age groups | The speed of elementary school students is faster than that of college students |
| 37 | Fu2021 [32] | 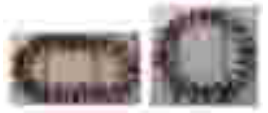 | Pedestrians | China | Outdoor | Oval, circle | Study the effect of setup geometry (oval, circular) on movement characteristics | There is an influence of curvature on the flow properties |
| 38 | Lu2021 [33] | 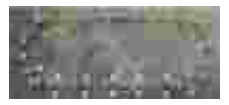 | Pedestrians | China | Outdoor | Oval | Analysis movement characteristics while maintaining social distancing | Social distancing measures cause pedestrians to maintain greater distances compared to normal walking |
| 39 | Ma2021 [34] | 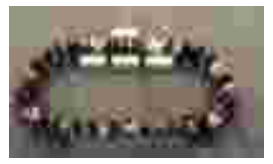 | Pedestrians | China | Outdoor | Oval | Study the synchronization of both legs in interacting pedestrians | Synchronization begins when the distance between pedestrians is too small for them to move forward |
| 40 | Wang2021 [35] | 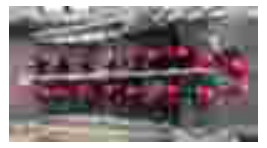 | Pedestrians | China | Indoor | Stairs | Investigate the effect of stair configuration on pedestrian ascent and descent | The analysis shows a unified behavioral mechanism for pedestrian single-file movement on both horizontal planes and stairs |
| 41 | Ye2021 [36] | 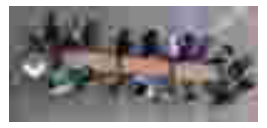 | Pedestrians | China | Outdoor | Stairs | Analysis of pedestrian movement on stairs with different motivations (normal and fast walking) | In fast walking condition, pedestrians move at higher velocities compared to normal walking |
| 42 | Ciuffo2021 [51] | 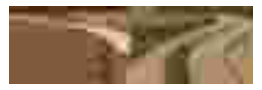 | Cars (autonomous vehicle) | Hungary | Outdoor | Motorway with random shape | Evaluate the impact of adaptive cruise control systems on vehicle flow | Adaptive cruise control systems tend to increase traffic instability |
| 43 | Lian2022 [37] | 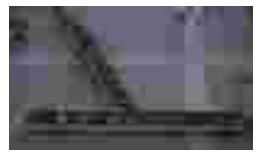 | Pedestrians | China | Outdoor | Branch | Investigate the merging flows of pedestrians at different angles | The speed-density relationship is not affected by the merging angle |
| 44 | Thompson2022 [38] | 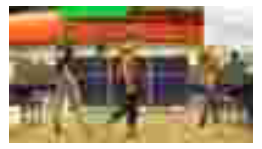 | Pedestrians | Sweden | Indoor | Oval | Analyze the step extent and contact buffer of pedestrians while walking | Step extent and contact buffer are key parameters for determining interpersonal spacing |
| 45 | Paetzke2023 [39] | 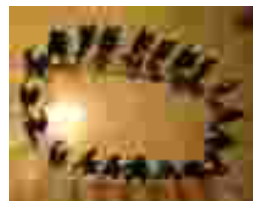 | Pedestrians | Germany | Indoor | Oval | Compare the movement of pedestrians in gender-homogeneous and gender-heterogeneous groups | No effect of the gender of neighboring pedestrians on movement |
| 46 | Li2023 [40] | 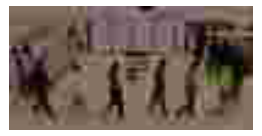 | Pedestrians | China | Indoor | Square with four straight corridors and four arcs | Investigate the influence of rhythm on male pedestrian movement | Rhythm affects movement and increases the frequency of stop-and-go waves |
| 47 | Bilintoh2023 [41] | 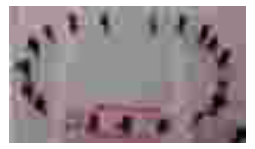 | Pedestrians | African students living in China | Outdoor | Oval | Investigate the movement characteristics of pedestrians from the same country living in different locations | Location influences movement |
| 48 | Li2024 [52] | 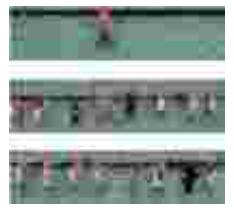 | Pedestrians | China | Indoor | One-dimensional observation area | Investigate the speed of pedestrians walking in specific water depths | Speed increases as the water depth increases |

* The main contribution and result reported as they appear in the original paper by the authors.

## B. Data Collection

**Table 4** Overview of data collection from experiments on single-file pedestrian movement during ground-level evacuation scenarios with closed boundaries, as reviewed by the authors.

| | Experiment (AuthorYear [cite]) | Country/territory (where the experiment performed) | Surrounding environment | Setup shape/type | Data type | Data collection device | Data collection process |
|---|---|---|---|---|---|---|---|
| 1 | Seyfried2005 [1] | Germany | Indoor | Oval | Time instances | Digital camera (side-view, video recordings) | Semi-automatic |
| | | | | | Head trajectories | Stereo vision camera (bird's-eye view) | Automatic |
| 2 | Chattaraj2009 [2] | India | Outdoor | Oval | Time instances | Digital camera (side-view, video recordings) | Semi-automatic |
| 3 | Lukowski2009 [5] | Germany | Indoor | Oval | Time instances | Digital camera (side-view, video recordings) | Semi-automatic |
| | | | | | Head trajectories | | |
| 4 | Liu2009 [6] | China | Outdoor | Oval | Head trajectories | Digital camera (bird's-eye view, video recording) | Automatic |
| 5 | Jezbera2010 [7] | Czech Republic | Outdoor | Circle | Time instances | Light gate | Automatic |
| 6 | Seyfried2010 [8,58] | Germany | Indoor | Oval | Head trajectories | Digital camera (bird's-eye view, video recording) | Automatic |
| 7 | Yanagisawa2012 [9] | Japan | Indoor | Circle | Time instances | No details | No details |
| 8 | Jelic2012 [10,54] | France | Indoor | Circle | Head and shoulders trajectories | Infrared camera (VICON motion capture system) | Automatic |
| 9 | Song2013 [11] | China | Outdoor | Oval | Head trajectories | Digital camera (bird's-eye view, video recording) | Automatic |
| 10 | Cao2016 [12] | China | Outdoor | Oval | Head trajectories | Digital camera (bird's-eye view, video recording) | Automatic |
| 11 | Ziemer2016 [13] | Germany | Indoor | Oval | Head trajectories | Digital camera (bird's-eye view, video recording) | Automatic |
| 12 | Zhao2017 [14] | China | Outdoor | Circle | Head trajectories | Digital camera (bird's-eye view, video recording) | Automatic |
| 13 | Ikeda2017 [16] | Japan | Indoor | Circle | Frontopolar/ brain activity signals | NIRS | Automatic |
| | | | | | Number of pedestrians | Digital camera (side-view around the circle, video recording) | Semi-automatic |
| 14 | Gulhare2018 [17] | India | Outdoor | Oval | Time instances | Digital camera (side-view, video recordings) | Semi-automatic |
| 15 | Ma2018 [19] | China | Outdoor | Oval | Head and foot trajectories | Digital camera (bird's-eye view, video recordings) | Automatic (mean-shift algorithm) |
| 16 | Wang2018 [21] | Germany | Indoor | Oval | Head trajectories | Digital camera (bird's-eye view, video recordings) | Automatic |
| 17 | Appert-Rolland2018 [22] | France | Indoor | Circle (virtual, without predefined path) | Head trajectories | Infrared camera (VICON motion capture system) | Automatic |
| 18 | Cao2019 [23] | China | Outdoor | Rectangle with four straight corridors and four arcs | Head trajectories | Digital camera (bird's-eye view, video recordings) | Automatic |
| 19 | Jin2019 [24, 55] | China | Outdoor | Circle | Head trajectories | UAV drone camera (bird's-eye view, video recordings) | Automatic |
| 20 | Ren2019 [25] | China | Outdoor | Oval | Head trajectories | Digital camera (bird's-eye view, video recordings) | Automatic |
| 21 | Subaih2019 [26, 56] | Palestine | Indoor | Oval | Head trajectories | Digital camera (side-view, video recordings) | Semi-automatic |
| 22 | Zeng2019 [28,60,61] | China | Outdoor | Oval | Head trajectories | Digital camera (bird's-eye view, video recordings) | Automatic |
| 23 | Ma2020 [29] | China | Outdoor | Oval | Head trajectories | Digital camera (bird's-eye view, video recordings) | Automatic |
| 24 | Wang2020a [30] | China | Indoor | Rectangle | Head trajectories | Digital camera (bird's-eye view, video recordings) | Automatic |
| 25 | Fu2021 [32] | China | Outdoor | Oval, circle | Head trajectories | Digital camera (bird's-eye view, video recordings) | Automatic |
| 26 | Lu2021 [33] | China | Outdoor | Oval | Head trajectories | Digital camera (bird's-eye view, video recordings) | Automatic |
| 27 | Ma2021 [34] | China | Outdoor | Oval | Head and foot trajectories | Camcorders | Automatic |
| 28 | Lian2022 [37] | China | Outdoor | Branch | Head trajectories | Digital camera (bird's-eye view, video recordings) | Automatic |
| 29 | Thompson2022 [38] | Sweden | Indoor | Oval | Right shoulder, hip, knee, the tip of the toe, and heel trajectories | Digital camera (side-view, video recordings) | Automatic |
| 30 | Paetzke2023 [39] | Germany | Indoor | Oval | Head trajectories | Digital camera (bird's-eye view, video recordings) | Automatic |
| 31 | Li2023 [40] | China | Indoor | Square with four straight corridors and four arcs | Head trajectories | UWB | Automatic |
| 31 | Li2023 [40] | China | Indoor | Square with four straight corridors and four arcs | No details | DJI camera (side-view, video recordings, entire setup) | No details |
| 32 | Bilintoh2023 [41] | African students living in China | Outdoor | Oval | Time instances | Digital camera (bird's-eye view, video recordings) | Semi-automatic |

# C. Collected experimental data for testing the proposed analysis tool

**Table 5** The details of the experiments, which are used to test our proposed analysis tool, comprise 28 datasets.

| | Experiment (AuthorYear [cite]) | Dimensions of the experimental setup: Investigates | Setup Central circumference [m] | Straight part length [m] | Measurement area length [m]) | Corridor width (straight, curved) [m] | Radius [m] | Trajectory extraction | Camera top/side view | Frame-rate (fps) |
|---|---|---|---|---|---|---|---|---|---|---|
| 1 | Lukowski2009 [5] | Motivation - haste | 17.30 | 4 | 2 | 0.8, 1.2 | 2.20 | Manually | Side view | 25 |
| 2 | Seyfried2010 [58] | Stop-and-go waves | 26.80 | 4 | 4 | 0.7, 0.7 | 3.00 | PeTrack | Top view | 25 |
| 3 | Cao2016 [12] | Age | 25.70 | 5 | - | 0.8, 0.8 | 2.90 | PeTrack | Top view | 25 |
| 4 | Ziemer2016 [13] | Congestion Dynamics | 26.84 | 4 | - | 1.0, 1.0 | 3.00 | PeTrack | Top view | 16 |
| 5 | Wang2018 [21] | Step style | 16.6 | 2.5 | - | 0.8, 0.8 | 2.25 | PeTrack | Top view | 25 |
| 6 | Subaih2019 [26] | Gender | 17.27 | 3.14 | 3.14 | 0.6, 0.6 | 2.05 | PeTrack | Side view | 25 |
| 7 | Ren2019 [25] | Age | 25.70 | 5 | - | 0.8, 0.8 | 2.50 | PeTrack | Side view | 25 |
| 8 | Zeng2019 [28] | Motivation - music | 21.93 | 5 | - | 0.8, 0.8 | 1.90 | PeTrack | Top view | 25 |
| 9 | Ma2020 [29] | Height constraints | 28.08 | 4 | 3 | 0.8, 0.8 | 2.4 | PeTrack | Top view | 25 |
| 10 | Paetzke2023 [39] | Gender | 14.97 | 2.3 | - | 0.8, 0.8 | 1.65 | PeTrack | Top view | 25 |